\documentclass[12pt]{article}
\pdfoutput=1
\usepackage{putex}
\usepackage{feyn}
\usepackage[vcentermath]{youngtab}
\usepackage{subfig}
\usepackage{lscape}

\usepackage{graphicx}
\usepackage{epstopdf}
\usepackage{enumerate}
\usepackage{cite}
\usepackage{tensor}
\usepackage{slashed}
\usepackage{amsmath}
\usepackage{amssymb}
\usepackage{mathrsfs}
\usepackage{lgrind}
\usepackage{comment}

\usepackage{bbm}

\usepackage{hyperref}

\numberwithin{equation}{section}

\newcommand {\be} {\begin {equation}}
\newcommand {\ee} {\end {equation}}

\newcommand {\bes} {\begin {equation*}}
\newcommand {\ees} {\end {equation*}}

\newcommand{\Lagr}{\mathcal{L}}

\newcommand{\eps}{\epsilon}

\def\CH{{\cal H}}

\newcommand{\beq}{\begin{equation}}
\newcommand{\eeq}{\end{equation}}

\def\be{ \begin{equation} }
\def\ee{ \end{equation} }

\def\Tr{{\textrm{Tr}}}
\begin{document}

\preprint{PUPT-2504}

\institution{PU}{Department of Physics, Princeton University, Princeton, NJ 08544}
\institution{PCTS}{Princeton Center for Theoretical Science, Princeton University, Princeton, NJ 08544}

\title{
Yukawa CFTs and Emergent Supersymmetry
}

\authors{Lin Fei,\worksat{\PU} Simone Giombi,\worksat{\PU} Igor R.~Klebanov\worksat{\PU,\PCTS} and Grigory Tarnopolsky\worksat{\PU}
}

\abstract{We study conformal field theories with Yukawa interactions in dimensions between 2 and 4; they provide UV completions of the 
Nambu-Jona-Lasinio and Gross-Neveu models which have four-fermion interactions. 
We compute the sphere free energy and certain operator scaling dimensions using dimensional continuation.
In the Gross-Neveu CFT with $N$ fermion degrees of freedom we obtain the first few terms in the $4-\eps$ expansion using the Gross-Neveu-Yukawa model, and the first few terms in the $2+\eps$ expansion using the four-fermion interaction. We then apply Pad\' e approximants to produce estimates in $d=3$. 
For $N=1$, which corresponds to one 2-component Majorana fermion, it has been suggested that the
Yukawa theory flows to a ${\cal N}=1$ supersymmetric CFT. We provide new evidence that the
$4-\eps$ expansion of the $N=1$ Gross-Neveu-Yukawa model respects the supersymmetry. 
Our extrapolations to $d=3$ appear to be in good agreement with the available results obtained using the numerical conformal bootstrap.
Continuation of this CFT to $d=2$ provides evidence that the Yukawa theory flows to the tri-critical Ising model.
We apply a similar approach to calculate the sphere free energy and operator scaling dimensions in the Nambu-Jona-Lasinio-Yukawa model, which has an additional 
$U(1)$ global symmetry. For $N=2$, which corresponds to one 2-component Dirac fermion, this theory has an
emergent supersymmetry with 4 supercharges, and we provide new evidence for this.}

\date{}
\maketitle

\tableofcontents

\section{Introduction and Summary}

Physical applications of relativistic quantum field theories with four-fermion interactions date back to Fermi's theory of beta decay. 
The first application to strong interactions was the seminal Nambu and Jona-Lasinio (NJL) model.
In the original paper \cite{Nambu:1961tp} they considered the model in $3+1$ dimensions with a single 4-component Dirac fermion and Lagrangian
\begin{align}
{\cal L}_{\textrm{NJL}}= \bar{\psi}  {\not\,}\partial \psi  + \frac {g}{2} \left ( (\bar{\psi} \psi)^{2}- (\bar{\psi} \gamma_5\psi)^{2}\right ) \,
.
\label{NJLorig}
\end{align}
In addition to the $U(1)$ symmetry $\psi \rightarrow e^{i\beta }\psi$, 
this Lagrangian possesses the $U(1)$ chiral symmetry under $\psi \rightarrow e^{i\alpha \gamma_5}\psi$. Using the gap equation it was shown that the chiral $U(1)$ can be broken spontaneously,
giving rise to the massless Nambu-Goldstone boson. This was the discovery of the crucial role of chiral symmetry breaking in the physics of strong interactions.

One of the goals of this paper is to study a generalization of (\ref{NJLorig}) to $N_f$ 4-component Dirac fermions $\psi_j$, $j=1, \ldots N_f$: 
\begin{align}
{\cal L}_{\textrm{NJL}}=  \bar{\psi}_{j}  {\not\,}\partial \psi^{j} +\frac{ g}{2} \left ( (\bar{\psi}_{j}\psi^{j})^{2}- (\bar{\psi}_{j}\gamma_5\psi^{j})^{2}\right ) \,
,
\label{NJLlag}
\end{align}
and its continuation to dimensions below $4$.
We define $N=4 N_f$, so that $N$ is the number of 2-component Majorana fermions in $d=3$.
In addition to the chiral $U(1)$ symmetry $\psi_j \rightarrow e^{i\alpha \gamma_5}\psi_j$ this multi-flavor NJL model possesses a $U(N_f)$ symmetry.\footnote{
It is also often called the chiral Gross-Neveu model \cite{Gross:1974jv}; in $d=2$
it is equivalent to the
$SU(2 N_f)$ Thirring model \cite{Witten:1978qu, Bondi:1989nq}.}
When considered in $2<d<4$ this model gives rise to an interacting conformal field theory which describes the second-order phase transition separating the phases where the
$U(1)$ chiral symmetry is broken and restored.

In addition to studying the NJL model with the $U(1)$ chiral symmetry, we will present new results for the Gross-Neveu (GN) model \cite{Gross:1974jv}, which  has a simpler quartic interaction
\begin{align}
{\cal L}_{\textrm{GN}}= \bar{\psi}_{j}  {\not\,}\partial \psi^{j}+\frac{g}{2} (\bar{\psi}_{j}\psi^{j})^{2}  \,
.
\label{GNlag}
\end{align}
Instead of the continuous chiral symmetry it possesses the discrete chiral symmetry $\psi_j \rightarrow \gamma_5 \psi_j$. As discovered in \cite{Gross:1974jv},
in $d=2$ this theory is asymptotically free for $N>2$. When considered in $2<d<4$ this is believed to be an interacting conformal field theory which describes the second-order phase transition where the
discrete chiral symmetry is broken.

In $d>2$ the four-fermion interactions (\ref{NJLlag}) and (\ref{GNlag}) are non-renormalizable. While they are renormalizable in the sense of the $1/N$ expansion \cite{Gross:1975vu},
at finite $N$ it is important to know the UV completion of these theories. In \cite{Hasenfratz:1991it,ZinnJustin:1991yn} it was suggested that the UV completion in $2<d<4$
is provided by the appropriate Yukawa theories. The UV completion of the GN model (\ref{GNlag}) contains a real scalar field $\sigma$ 
and $N_f$ 4-component Dirac fermions $\psi_j$, $j=1, \ldots N_f$:
\begin{align}
\Lagr_{\textrm{GNY}} =  \frac{1}{2}(\partial_{\mu}\sigma)^{2}+\bar{\psi}_{j}{\not\,}\partial \psi^{j} + g_{1}\sigma \bar{\psi}_{j}\psi^{j}+\frac{1}{24}g_{2}\sigma^{4}\ .
\label{GNY-Lag}
\end{align} 
 This theory,  known as the
Gross-Neveu-Yukawa (GNY) model, will be discussed in section \ref{GNYsection}. 

The UV completion of the $U(1)$ symmetric NJL model (\ref{NJLlag}), which contains a
complex scalar field $\phi=\phi_1+i\phi_2$, was introduced in \cite{ZinnJustin:1991yn}: 
\begin{equation}
\Lagr_{\textrm{NJLY}} = \frac{1}{2}(\partial_{\mu}\phi_1)^2+\frac{1}{2}(\partial_{\mu}\phi_2)^2+\bar{\psi_j}\slashed{\partial}\psi^j +
g_1\bar{\psi_j}(\phi_1+i\gamma_5 \phi_2)\psi^j+\frac{1}{24}g_2(\phi \bar \phi)^2\ . \label{NJLYlagrange}
\end{equation}
It has a continuous $U(1)$ chiral symmetry under\footnote{In $d=3$, one may express the Lagrangian (\ref{NJLlag}) in terms 
of $2N_f$ 2-component Dirac spinors $\chi^i, \chi^{i+N_f}$ by writing $\psi^i = (\chi^i, \chi^{i+N_f})$, $i=1,\ldots, N_f$. See for instance \cite{Kubota:2001kk} for the explicit relation between 4-component and 2-component notations in 3d. The 2-component spinors  $\chi^{i\pm} = (\chi^i\pm \chi^{i+N_f})/\sqrt{2}$ 
have charge $\pm 1$ under the $U(1)$ symmetry (\ref{U1-A}).}
\begin{equation} 
\psi_j \rightarrow e^{i\alpha \gamma_5} \psi_j\ , \qquad \phi \rightarrow e^{-2 i\alpha} \phi\,.
\label{U1-A}
\end{equation}
This theory, which we will call the Nambu-Jona-Lasinio-Yukawa model (NJLY), will be discussed in section \ref{NJLYsection}.

There is a large body of literature on the GN and NJL CFTs in $d=3$ and their applications; see, for example, \cite{Gracey:1990wi,Hasenfratz:1991it,ZinnJustin:1991yn,Gracey:1993cq,Hands:1992be,Rosenstein:1993zf,Roy:2012wz,Karkkainen:1993ef,Hands:2001cs,Moshe:2003xn,Christofi:2006zt,Chandrasekharan:2013aya,Iliesiu:2015qra,Diab:2016spb}.
 We will carry out further studies of these CFTs using the $4-\eps$ and $2+\eps$ expansions followed by Pad\' e extrapolations. 
In addition to studying the scaling dimensions of some low-lying operators, we will calculate the sphere free energy $F$.
The latter determines the universal entanglement entropy across a circle \cite{Casini:2011kv}, and is the quantity that enters the $F$-theorem
\cite{Myers:2010xs,Jafferis:2011zi,Klebanov:2011gs,Casini:2012ei}. We will also discuss $C_T$, the normalization of the correlation function of two stress-energy tensors.
For the GN model, the $1/N$ and $\epsilon$ expansions of $C_T$ were studied in \cite{Diab:2016spb}; in this paper we extend these results to the NJL model and also provide
the numerical estimates in $d=3$ for various values of $N$. 

\begin{figure}[h!]
    \centering
    \subfloat{{\includegraphics[width=7.5cm]{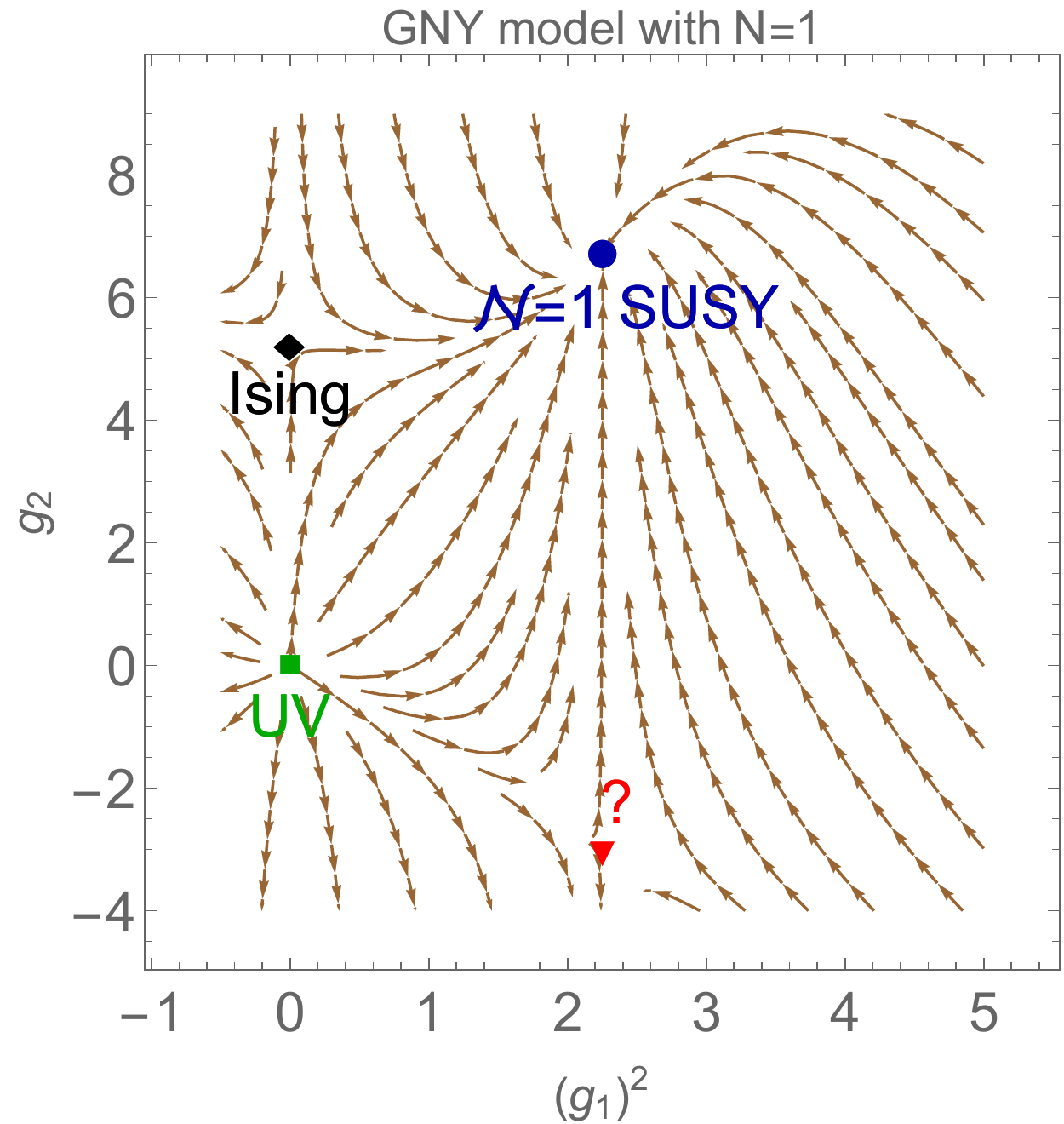} }}
    \qquad
    \subfloat{{\includegraphics[width=7.5cm]{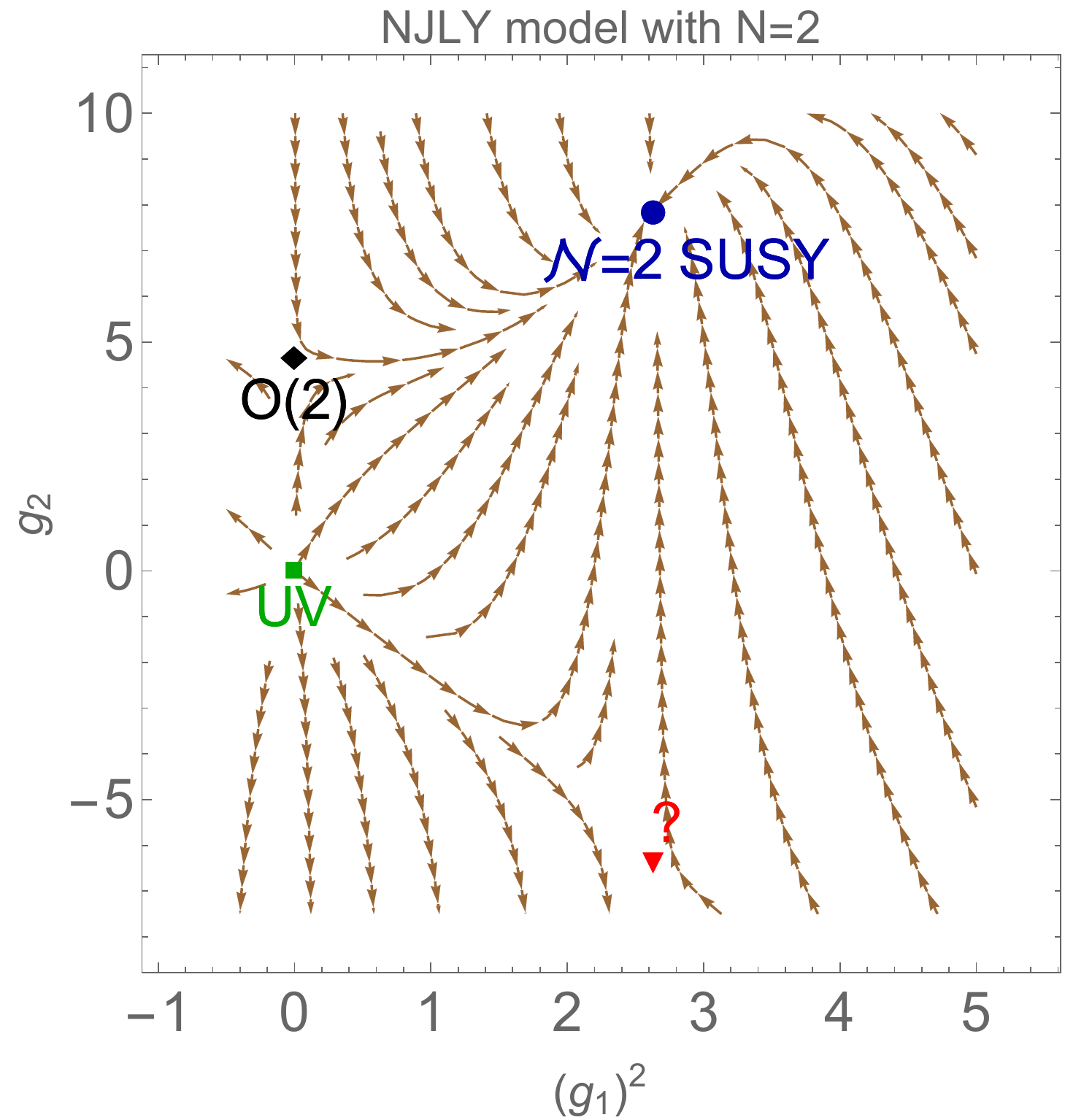} }}
    \caption{RG flow and fixed point structure for the GNY model with $N=1$ (one Majorana fermion in $d=3$) and the NJLY model with $N=2$ (one 
Dirac fermion in $d=3$), obtained from the one-loop $\beta$-functions (\ref{betasGNY}) and (\ref{NJLYbeta}) in $d=4-\eps$. The attractive IR fixed points have ``emergent" 
supersymmetry with 2 and 4 supercharges respectively. The red triangles denote unstable fixed points with negative quartic potential which 
can be seen in the one-loop analysis in $d=4-\eps$; their fate in $d=3$ is unclear.}
    \label{RGs}
\end{figure}

When considered for $N_f=1/2$, i.e. the single 4-component Majorana fermion (which is equivalent to one Dirac fermion in $d=3$), 
the NJLY model is expected to flow to the well-known supersymmetric Wess-Zumino model with 4 supercharges.
In $d<4$ this theory defines a CFT with ``emergent supersymmetry" \cite{Thomas,Lee:2006if}, in the sense 
that the RG flow drives the interactions to a supersymmetric IR stable fixed point, where the global $U(1)$ symmetry becomes the $U(1)_R$
symmetry (see figure \ref{RGs}). We will provide additional 
evidence for this using the $4-\epsilon$ expansion of the NJLY model with $N_f=1/2$ to two loops, both for certain scaling dimensions and for the sphere free energy. A three loop calculation of scaling dimensions, which supports the emergent supersymmetry, was carried out recently \cite{Zerf:2016fti}.  
 
Even more intriguingly, when the GNY model is continued to $N_f=1/4$, which corresponds to a single 2-component Majorana fermion in $d=3$, it appears to flow to a CFT with 2 supercharges 
\cite{Thomas,Grover:2013rc,Bashkirov:2013vya,Iliesiu:2015qra,Shimada:2015gda}. 
We will show that the $\mathcal{O}(\eps^2)$ corrections to scaling dimensions of operators continue to respect this emergent supersymmetry.\footnote{This is reminiscent of the symmetry enhancement from $Sp(2)$ to the supergroup $OSp(1|2)$ at the IR stable fixed point of a cubic theory in $6-\eps$ dimensions \cite{Fei:2015kta}.} This provides new support
for the existence of an ${\cal N}=1$ supersymmetric CFT in $d=3$. Our Pad\' e extrapolations of operator dimensions including the $\eps^2$ corrections
are in good agreement with the conformal bootstrap approach to the ${\cal N}=1$ supersymmetric CFT in $d=3$ \cite{Iliesiu:2015qra}. 
We also estimate $C_T$ and $F$ for this theory.
These results will be presented in section \ref{emergentSUSY}.

\section{The Gross-Neveu-Yukawa Model}

\label{GNYsection}

The $\beta$-functions for the GNY model with action (\ref{GNY-Lag}) in $d=4-\eps$, up to two-loop order, are  \cite{Karkkainen:1993ef}\footnote{We have reproduced
them using the general two-loop results \cite{Machacek:1983tz, Machacek:1983fi, Machacek:1984zw,Jack:1990eb,Pernici:1999nw} for the Yukawa theories, which are reviewed in the
 Appendix.}
\begin{align}
&\beta_{g_{2}}= -\epsilon g_{2} + \frac{1}{(4\pi)^{2}} \Big(3g_{2}^{2}+2N g_{1}^{2}g_{2}-12N g_{1}^{4}\Big)+\frac{1}{(4 \pi )^4}
\Big(96 N g_{1}^6 +7 N g_{1}^4 g_{2} -3N g_{1}^2 g_{2}^2 -\frac{17 g_{2}^3}{3}\Big)\,, \notag\\
&\beta_{g_{1}} = -\frac{\epsilon}{2}g_{1}+\frac{N+6}{2(4\pi)^{2}}g_{1}^{3}+\frac{1}{(4 \pi )^4}\Big(-\frac{3}{4} (4 N+3)g_{1}^{5} -2 g_{1}^3 g_{2}+\frac{g_{1}g_{2}^2}{12}\Big)\,,\label{betasGNY}
\end{align}
where   $N = N_f{\rm tr}\textbf{1}=4N_f$.
The model possesses an IR stable fixed point at the critical couplings $g_i^*$ given by
\begin{align}
&\frac {(g_{1}^{*})^{2}} {(4\pi)^{2}}=\frac{1}{N+6}\eps +\frac{ (N+66)\sqrt{N^2+132 N+36}-N^2+516 N+882}{108 (N+6)^3}\eps^2+ \mathcal{O}(\eps^3)\,, \notag\\
& \frac{g_{2}^{*}} { (4\pi)^{2}}= \frac{-N+6+\sqrt{N^2+132N+36}}{6(N+6)}\eps +\frac{1}{54(N+6)^3\sqrt{N^2+132N+36}} \notag\\
&~~~~~~~~~~~\times \Big(3N^4+155N^3+2745N^2-2538N+7344\notag\\
&~~~~~~~~~~~~~~~~~~-(3N^3-43N^2-1545N-1224) \sqrt{N^2+132 N+36}\Big)\eps^2 +\mathcal{O}(\eps^3)\,. \label{GNYcoupl}
\end{align}
Of course, there is also a fixed point $g_1^*=0$, $g_2^*=g_2^{\rm Ising}=\frac{16\pi^2\eps}{3}+\mathcal{O}(\eps^2)$ which corresponds to the decoupled product 
of the single-scalar Wilson-Fisher fixed point and $N_f$ free fermions. By looking at the derivative of the beta functions at the fixed points, one can 
verify that (\ref{GNYcoupl}) is attractive for all $N_f$, so one can flow to it from the ``Ising" fixed point along a relevant direction. Let us mention 
that there is formally also a third fixed point obtained from (\ref{GNYcoupl}) by changing the sign of  $\sqrt{N^2+132 N+36}$.
This fixed point is unstable in $d=4-\eps$ due to the negative 
quartic coupling, $g_2^* <0 $, but its dimensional continuation may produce a CFT in $d=3$. 

The scaling dimensions of $\sigma$ and $\psi$ are found to be \cite{Karkkainen:1993ef}
\begin{align}
\Delta_\sigma &= 1-\frac{\eps}{2} + \frac{1}{2 (4\pi)^2} Ng_1^2+ \frac{1}{(4\pi)^4}\left (\frac{1}{12} g_2^2 
-\frac{5}{4}N g_1^4 \right )\,, \notag \\
\Delta_\psi&= \frac{3-\eps}{2} +\frac{1}{2(4\pi)^2}g_1^2 
-\frac{1}{8 (4\pi)^4} (3N +1) g_1^4 
\ .\label{GNYdim}
\end{align}
At the IR stable fixed point (\ref{GNYcoupl}), one gets\footnote{Throughout the paper, one can obtain the corresponding $\epsilon$ expansions at the unstable fixed point 
by changing the sign of the square root.}
\begin{equation}
\begin{aligned}
\Delta_{\sigma} &= 1-\frac{3}{N+6}\eps + \frac{52N^2-57N+36+(11N+6)\sqrt{N^2+132N+36}}{36(N+6)^3}\eps^2+ \mathcal{O}(\eps^3)\,, \\
\Delta_{\psi} &= \frac{3}{2}  -\frac{N+5}{2(N+6)}\eps + \frac{-82N^2+3N+720+(N+66)\sqrt{N^2+132N+36}}{216(N+6)^3}\eps^2+\mathcal{O}(\eps^3)\,.
\label{GNYdims}
\end{aligned}
\end{equation} 
These dimensions agree with \cite{Rosenstein:1993zf} after correcting some typos in eq. (11) of that paper (in particular, the coefficient $33$ should be changed to $3$).
Our $\mathcal{O}(\eps)$ term in $\Delta_\psi$ corrects a typo in eq. (4.44) of \cite{Moshe:2003xn}.

Setting $g_1^*=0, g_2^* = \frac{16\pi^2\eps}{3}+\mathcal{O}(\eps^2)$ in (\ref{GNYdim}), we may also recover the result at the Ising fixed point 
$\Delta_{\sigma}^{\rm Ising} = 1-\frac{\eps}{2}+\frac{\eps^2}{108}+\mathcal{O}(\eps^3)$. One can then see that the Yukawa operator $\sigma \bar\psi \psi$ 
is relevant at this decoupled fixed point, and can trigger a flow to the IR stable fixed point (\ref{GNYcoupl}). We expect this to be true in $d=3$ as well, 
since it is known that $\Delta_{\sigma}^{\rm 3d~Ising}\approx 0.518$ \cite{El-Showk:2014dwa}, and so $\Delta_{\sigma\bar\psi\psi}^{\rm 3d~Ising} \approx 2.518 <3$. 
 
The anomalous dimension of the operator $\sigma^2$, which determines the critical exponent $\nu^{-1}= 2- \gamma_{\sigma^2}$, may be read off from eq. (18) of 
\cite{Karkkainen:1993ef}
\begin{align}
 \gamma_{\sigma^2}= \frac{g_{2}}{(4\pi)^{2}} - \frac{1}{(4\pi)^{4}}(g_{2}^{2}+ N g_{2}g^{2}_{1}-2Ng_{1}^{4})+2 \gamma_{\sigma}\,.
 \label{GNYsigma2}
\end{align}
At the fixed point (\ref{GNYcoupl}) we find
\begin{align}
&\Delta_{\sigma^2} = d-2 + \gamma_{\sigma^2}= 2+ \frac{\sqrt{N^2+132N+36}- N -30}{6 (N+6)}\eps +\frac{1} { 54 (N+6)^3 \sqrt{N^2+132N+36}}\notag \\ 
&   ~~~~~~~~~~\times \Big((3 N^3+109 N^2 + 510 N + 684) \sqrt{N^2+132N+36} \notag\\
&~~~~~~~~~~~~~~~~~~- 3 N^4 - 658 N^3-333 N^2 -15174 N + 4104\Big)\eps^2 +  \mathcal{O}(\eps^3)\, .
\label{GNYsigmasq}
\end{align}
We have also calculated the one-loop anomalous dimensions of the operators $\bar{\psi}\psi$ and $\sigma^{3}$:
\begin{align}
&\Delta_{\bar{\psi}\psi} = \frac{(N+2)g_{1}^{2}}{(4\pi)^{2}}+2\Delta_{\psi}\,, \quad \Delta_{\sigma^{3}}= \frac{3g_{2}}{(4\pi)^{2}}+3\Delta_{\sigma}\,.
\label{GNYpsipsisigma3}
\end{align}
At higher orders these operator will mix, and one has to find the eigenvalues of their mixing matrix.  At the fixed point (\ref{GNYcoupl}), we find
 \begin{align}
&\Delta_{\bar{\psi}\psi} = 3-\frac{3}{N+6}\eps +\mathcal{O}(\eps^{2})\,, \quad \Delta_{\sigma^{3}}= 3+\frac{\sqrt{N^{2}+132 N+36}-N-12}{2(N+6)}\eps+\mathcal{O}(\eps^{2}) \,.
\label{GNYpsipsisigma3ep}
\end{align}
The first of these dimensions corresponds to a descendant of $\sigma$, as can be seen from the fact that it equals $2+\Delta_\sigma$.

Let us also review the known result for the $4-\eps$ expansion of $C_T$ in the GNY model, 
which was discussed in \cite{Diab:2016spb}; the diagrams contributing to the term $\sim g_1^2$ are shown in fig. 4.7 there.
Evaluation of these diagrams yields
\begin{align}
C_T  = N C_{T,f}+ C_{T,s}- \frac {5 N (g_1^*)^2}{12 (4 \pi)^2} 
= \frac{d}{S_d^2}\left(\frac{N}{2} + \frac{1}{d-1} - \frac{5N\epsilon}{12(N+6)}\right)\, ,
\label{CT-GNY}
\end{align}
where $S_d = 2\pi^{d/2}/\Gamma(d/2)$,
and we used the values of $C_T$ for free scalar and fermion theories:
\begin{align}
C_{T,s} = \frac{d}{(d-1) S_d^2}\ , \qquad C_{T,f}= \frac{d}{2 S_d^2}
\ .
\end{align}

\subsection{Free energy on $S^{4-\eps}$}

In order to renormalize the theory on curved space, one should add to the action all the relevant curvature couplings that are marginal 
in $d=4-\eps$ \cite{Brown:1980qq,Hathrell:1981zb}
\begin{equation}
\begin{aligned}
S_{\textrm{GNY}} = \int d^{d}x \sqrt{g}\Big(&\bar{\psi}_{i}{\not\,}\partial \psi^{i} 
+ \frac{1}{2}(\partial_{\mu}\sigma)^{2}+\frac{d-2}{8(d-1)}{\cal R}\sigma^2
+g_{1,0}\sigma \bar{\psi}_{i}\psi^{i}+\frac{g_{2,0}}{24}\sigma^{4}\\
&+\frac{\eta_0}{2}{\cal R}\sigma^2+a_0 W^2+b_0 E+c_0 {\cal R}^2\Big)\,,
\end{aligned}
\end{equation}
where ${\cal R}$ is the scalar curvature, $W^2$ is the square of the Weyl tensor, and $E$ the Euler density
\begin{equation}
E=  \mathcal{R}_{\mu\nu\lambda\rho}  \mathcal{R}^{\mu\nu\lambda \rho} -4 \mathcal{R}_{\mu\nu} \mathcal{R}^{\mu\nu} + \mathcal{R}^{2}\,.
\end{equation}
The parameters $\eta_0, a_0, b_0, c_0$ are bare curvature couplings whose renormalization can be fixed order by order in perturbation 
theory. On a sphere, the Weyl square term drops out, and to the order we work below we will only need the renormalization of 
the Euler coupling $b_0$ (the ${\cal R}^2$ term and the renormalization of conformal coupling are expected to play a role at higher orders 
\cite{Brown:1980qq,Hathrell:1981zb,Jack:1990eb}). The corresponding beta function can be extracted from the results of \cite{Jack:1990eb, Jack:1984vj}, 
and we find
\begin{align}
&\beta_{b}= \epsilon b -\frac{11N/4+1}{360(4\pi)^2}-\frac{1}{(4\pi)^8}\frac{1}{36}\left(\frac{9}{32}N g_1^6 + \frac{3}{8}N^2 g_1^6 + \frac{1}{4}N g_1^4 g_2 
- \frac{1}{96}N g_1^2 g_2^2 \right)+\ldots\,,
\label{betabGNY}
\end{align}
where $b$ is the renormalized coupling, and its the relation to the bare one $b_0=\mu^{-\eps}(b-\frac{11N/4+1}{360(4\pi)^2\eps}+\ldots)$ 
can be inferred from the above beta function. The 
coupling independent term is related to the $a$-anomaly of the free fermions and scalar. 

The calculation of the sphere free energy now proceeds as in \cite{Fei:2015oha, Giombi:2015haa}. Keeping terms that contribute up to order $\epsilon^2$, 
we have
\begin{align}
&F
=N F_f+ F_s -\frac{1}{2}g_{1,0}^{2} \int d^{d}xd^{d}y \sqrt{g_{x}}\sqrt{g_{y}} \langle \sigma \bar{\psi} \psi(x) \sigma \bar{\psi}\psi(y)\rangle_{0}\notag\\
&~~~~~- \frac{1}{2!(4!)^{2}}g_{2,0}^{2}\int d^{d}xd^{d}y \sqrt{g_{x}}\sqrt{g_{y}} \langle \sigma^{4}(x)\sigma^{4}(y)\rangle_{0} \notag\\
&~~~~~-\frac{1}{4!}g_{1,0}^{4}
\int d^{d}xd^{d}yd^{d}zd^{d}w \sqrt{g_{x}}\sqrt{g_{y}}\sqrt{g_{z}}\sqrt{g_{w}}\langle \sigma \bar{\psi}\psi(x) \sigma\bar{\psi}\psi(y) \sigma\bar{\psi}\psi(z)\sigma\bar{\psi}\psi(w)\rangle_{0} + \delta F_b\, , \label{GNYFree}
\end{align}
where $\delta F_b = b_0 \int d^d x \sqrt{g} E$ is the contribution of the curvature term, and $F_f$, $F_s$ are the sphere free energies of free 
fermion and scalar, which can be found in \cite{Giombi:2014xxa}. Starting from the flat space propagators
\begin{align}
&\langle \sigma(x)\sigma(y)\rangle = C_{\phi} \frac{1}{|x-y|^{d-2}}\,, \quad \langle \psi_{i}(x)\bar{\psi}^{j}(y)\rangle
=\delta^{j}_{i}C_{\psi}\frac{\gamma^{\mu} (x-y)_{\mu}}{|x-y|^{d}}\,,
\label{flat-prop}
\end{align}
where  $C_{\phi} = \Gamma(\frac{d}{2}-1)/(4\pi^{\frac{d}{2}})$ and $C_{\psi }=\Gamma(\frac{d}{2})/(2\pi^{\frac{d}{2}})$,
and then Weyl transforming to the sphere, we find
\begin{align}
 &\int d^d x d^d y \sqrt{g_x}\sqrt{g_y}\langle \sigma \bar{\psi} \psi(x) \sigma \bar{\psi}\psi(y)\rangle_{0} 
= N C_{\phi}C_{\psi}^{2} \int d^d x d^d y \frac{\sqrt{g_x}\sqrt{g_y}}{s(x,y)^{3d-4}} = N C_{\phi}C_{\psi}^{2} I_2(\frac{3d}{2}-2)\,,\notag \\
 &\int d^d x d^d y \sqrt{g_x}\sqrt{g_y}\langle \sigma^{4}(x)\sigma^{4}(y)\rangle_{0} 
=4! C_{\phi}^{4}\int d^d x d^d y \frac{\sqrt{g_x}\sqrt{g_y}}{s(x,y)^{4(d-2)}} = 4! C_{\phi}^{4} I_2(2d-4)\,.
 \end{align}
Here $I_2(\Delta)$ denotes the integrated 2-point function of an operator of dimension $\Delta$, 
which is given by \cite{Drummond:1977dn, Cardy:1988cwa, Klebanov:2011gs}
\begin{equation}
I_{2}(\Delta)=\int \frac{d^{d}xd^{d}y\sqrt{g_{x}}\sqrt{g_{y}}}{s(x,y)^{2\Delta}}= (2 R)^{2 (d-\Delta )}\frac{2^{1-d} \pi ^{d+\frac{1}{2}} \Gamma \left(\frac{d}{2}-\Delta \right)}{\Gamma \left(\frac{d+1}{2}\right) \Gamma (d-\Delta )}\, .
\label{I2}
\end{equation}
For the 4-point function, we find
 \begin{align}
 & \langle \sigma \bar{\psi}\psi(x) \sigma\bar{\psi}\psi(y) \sigma\bar{\psi}\psi(z)\sigma\bar{\psi}\psi(w)\rangle_{0} =6N^{2}C_{\phi}^{2}C_{\psi}^{4}
 \frac{1}{s_{xy}^{2d-2}s_{zw}^{2d-2}s_{xw}^{d-2}s_{yz}^{d-2}} \notag\\
 &~~~~~~~~~~~~-3NC_{\phi}^{2}C_{\psi}^{4}\bigg(\frac{s_{xw}^{2}s_{yz}^{2}-s_{xz}^{2}s_{yw}^{2}+s_{xy}^{2}s_{zw}^{2}}{(s_{xy}s_{yz}s_{zw}s_{xw})^{d}}\bigg) \times \bigg(\frac{2}{(s_{xy}s_{zw})^{d-2}}+\frac{1}{(s_{xz}s_{yw})^{d-2}}\bigg)\,, \label{PropGNY}
 \end{align}
where we used a shorthand notation for the chordal distance $s_{xy}\equiv s(x,y)$.
The integral of this 4-point function over the sphere cannot be calculated explicitly, but one can find it as a series in $d=4-\epsilon$. For this we used the Mellin-Barnes approach, which is described in \cite{Fei:2015oha, Giombi:2015haa}. The result for the integral reads
\begin{align}
&\int d^{d}xd^{d}yd^{d}zd^{d}w \sqrt{g_{x}}\sqrt{g_{y}}\sqrt{g_{z}}\sqrt{g_{w}}\langle \sigma \bar{\psi}\psi(x) \sigma\bar{\psi}\psi(y) \sigma\bar{\psi}\psi(z)\sigma\bar{\psi}\psi(w)\rangle_{0} = \notag\\
&~~~~~~~~~~~~~~=-\frac{N (N+6)}{2 (4\pi) ^4 \epsilon }
-\frac{N \big(N-6+6 (N+6) (3+\gamma+\log (4\pi R^{2}))\big)}{12 (4\pi) ^4} +\mathcal{O}(\epsilon).
\end{align}
Putting everything together, we find for the free energy in $d=4-\epsilon$
\begin{align}
&F
=N F_f+ F_s -\frac{1}{2}g_{1,0}^{2} N C_{\phi}C_{\psi}^{2}  I_{2}\big(\frac{3}{2}d-2\big)- \frac{1}{2\cdot 4!}g_{2,0}^{2}C_{\phi}^{4}I_{2}(2d-4) \notag\\
&~~~~~~+\frac{1}{4!}g_{1,0}^{4}
\bigg(\frac{N (N+6)}{2 (4\pi) ^4 \epsilon }
+\frac{N \big(N-6+6 (N+6) (3+\gamma+\log (4\pi R^{2}))\big)}{12 (4\pi) ^4} + \mathcal{O}(\epsilon)\bigg) + \delta F_b\ .
\end{align}
Now replacing the bare couplings with the renormalized ones
\begin{align}
&g_{1,0} = \mu^{\frac{\epsilon}{2}}\Big(g_{1}+\frac{N+6}{32\pi^{2}}\frac{g_{1}^{3}}{\epsilon}+\dots\Big), \quad g_{2,0}= \mu^{\epsilon}\big(g_{2}+\dots \big),
\quad b_0=\mu^{-\eps}\Big(b-\frac{11N/4+1}{360(4\pi)^2\eps}+\ldots\Big)
\end{align}
we find that all pole cancels, and the free energy is a finite function of the renormalized couplings $g_1, g_2, b$.\footnote{Note that the 
coupling dependent part in the renormalization of $b$ is necessary to cancel poles coming from diagrams 
at the next order. However, we still have to carefully include the Euler term, as in \cite{Fei:2015oha,Giombi:2015haa}, 
as it affects the free energy at the fixed point to order $\eps^2$.}

As explained in \cite{Fei:2015oha}, in order to calculate the free energy at the critical point we should now tune all couplings, including $b$, to their 
fixed point values. Using (\ref{GNYcoupl}), we get
\begin{align}
F
&= N F_f+ F_s -\frac{N \epsilon }{48 (N+6)} \notag\\
&~~~-\frac{\left( \left(N^2+99 N+18\right)\sqrt{N^{2} +132N+36}+\left(80 N^2+2103 N+6381\right) N+108\right) \epsilon ^2}{7776 (N+6)^3} \notag \\
& ~~~ + \delta F_b + \mathcal{O}(\epsilon^{3})\,.
\label{withoutcurve}
\end{align}
From the curvature beta function (\ref{betabGNY}), we find at the critical point
\begin{equation}
b_*=\frac{11N/4+1}{360(4\pi)^2\eps}+\frac{1}{(4\pi)^8}\frac{1}{36}\left(\frac{9}{32}N (g_1^*)^6 + \frac{3}{8}N^2 (g_1^*)^6 
+ \frac{1}{4}N (g_1^*)^4 g_2^* - \frac{1}{96}N (g_1^*)^2 (g_2^*)^2 \right)\frac{1}{\eps}
\end{equation}
and, using (\ref{GNYcoupl}) and $\int d^dx \sqrt{g}E = 64\pi^2+\mathcal{O}(\eps)$, we find that the Euler term contributes
\begin{align}
\delta F_b =\frac{N(882+66\sqrt{N^2+132N+36}+N(516-N+\sqrt{N^2+132N+36}))}{15552(N+6)^3}\eps^2 +\mathcal{O}(\eps^3)\,.
\end{align}
Substituting this into (\ref{withoutcurve}), and writing the result in terms of $\tilde{F}= -\sin(\pi d/2)F$, we find
\begin{align}
&\tilde{F}= N \tilde F_f + \tilde F_s  -\frac{N \pi \epsilon ^2}{96 (N+6)} -\frac{1}{31104(N+6)^3}\Big(161 N^3 + 3690 N^2 + 11880N + 216\notag\\
&~~~~~~~~~~~+\left(N^2+132 N+36\right)\sqrt{N^{2} +132N+36} \Big) \pi\epsilon ^3+\mathcal{O}\big(\epsilon^{4}\big).
\label{tFGNYall}
\end{align}

\subsection{$2+\epsilon$ expansions}

In this section we review the known results for operator dimensions at the UV fixed point of the Gross-Neveu model in $d=2+\eps$, and then 
compute its sphere free energy to order $\eps^3$. 

The action for the Gross-Neveu model \cite{Gross:1974jv}  in Euclidean space in terms of bare fields and coupling reads
\begin{align}
S_{\textrm{GN}}&= -\int d^{d}x\sqrt{g}\Big(\bar{\psi}_{i} {\not\,}\partial \psi^{i}+\frac{1}{2}g_{0}(\bar{\psi}_{i}\psi^{i})^{2}\Big)+b_{0}\int d^{d}x \sqrt{g}\mathcal{R}\,
,
\end{align}
where $i=1,\ldots 2 N_f$, and we have included the Euler term which is needed for the calculation of the sphere free energy below. 

The $\beta$-function for the renormalized coupling constant $g$ in $d=2+\epsilon $ is known to be 
\cite{Kivel:1993wq, Moshe:2003xn,Gracey:2008mf}
\begin{align}
\beta =\epsilon g -\frac{N-2}{2\pi}g^{2}+\frac{N-2}{4\pi^{2}}g^{3}+\frac{ (N-2) (N-7)}{32 \pi ^3}g^4+\mathcal{O}(g^{5})\, .
\end{align}
Therefore, one can see that there is a perturbative $\textrm{UV}$ fixed point at a critical
coupling $g_{*}$ given by\footnote{This corrects a typo in $g_{*}$ in  \cite{Moshe:2003xn} on page 59 (there it is denoted by $u_{c}$).}
\begin{align}
\quad g_{*}= \frac{2\pi }{N-2}\epsilon+\frac{2 \pi }{(N-2)^2} \epsilon ^2+\frac{\pi  (N+1)}{2 (N-2)^3} \epsilon ^3+\mathcal{O}(\epsilon^{4})\,.
\label{GNcritcoup}
\end{align}
The scaling dimensions are found to be \cite{Gracey:2008mf}
\begin{align}
\Delta_\psi &= \frac{1+\eps}{2} + \frac{N-1}{8 \pi^2}  g^2- \frac{(N-1)(N-2)}{32 \pi^3} g^3 +
\mathcal{O}(g^{4})\ ,\notag \\
\Delta_\sigma &= 1+ \eps - \frac{N-1}{2\pi}  g + \frac{N-1}{8\pi^2} g^2 
+\frac{(N-1)(2N-3)}{32\pi^3} g^3  + \mathcal{O}(g^{4})
\ ,\label{GNdimraw}
\end{align}
where $\sigma\sim \bar \psi \psi$.
At the UV fixed point this gives
\begin{align}
\Delta_{\psi} &= \frac{1}{2} + \frac{1}{2}\eps+ \frac{N-1}{4 (N-2)^2} \eps^2  -\frac{(N-1) (N-6) }{8 (N-2)^3} \eps^3 + \mathcal{O}(\eps^4)\, ,\\ 
\Delta_{\sigma} &= 1-\frac{1}{N-2}\eps - \frac {N-1}{2 (N-2)^2} \eps^2 + \frac {N(N  -1) }{4 (N-2)^3} \eps^3  + \mathcal{O}(\eps^4)\, ,
\label{GNdim}
\end{align}
It is also not hard to determine the dimension 
\begin{align}
\Delta_{\sigma^2} = d+ \beta'(g_*) =  2+\frac{1}{N-2}\eps^2 + \frac{N-3}{2(N-2)^2}\eps^3 + 
\mathcal{O}(\eps^4)\ .
\end{align}

Let us now turn to the calculation of the free energy on $S^{2+\eps}$. To order $g^4$, we have
\begin{align}
F
=&N F_f -\frac{1}{2!}\Big(\frac{g_{0}}{2}\Big)^{2}S_{2}-
\frac{1}{3!}\Big(\frac{g_{0}}{2}\Big)^{3}S_{3}-
\frac{1}{4!}\Big(\frac{g_{0}}{2}\Big)^{4}S_{4}+b_{0}\int d^{d}x \sqrt{g}\mathcal{R}\,,
\end{align}
where 
$F_f$ is the free fermion contribution, derived as a function of $d$ in \cite{Giombi:2014xxa},
and
\begin{align}
S_{n} = \int \prod_{i=1}^{n}dx_{i}\sqrt{g_{x_{i}}} \langle \psi^{4}(x_{1})...\psi^{4}(x_{n})\rangle_{0}^{\textrm{conn}}\,,
\end{align}
with   $\psi^{4}\equiv (\bar{\psi}_{i}\psi^{i})^{2}$.
Using the flat space fermion propagator in (\ref{flat-prop}), and then performing a Weyl transformation to the sphere, we find
\begin{align}
&S_{2}= 2N(N-1)C_{\psi}^{4}I_{2}(2d-2) \,, \notag \\
&S_{3}=8N(N-1)(N-2) C_{\psi}^{6} I_{3}(2d-2)\,,
 \end{align}
where the integral $I_{2}(\Delta)$ is given in (\ref{I2}), and $I_{3}(\Delta)$ denotes the integrated 3-point function 
\cite{Drummond:1977dn, Cardy:1988cwa, Klebanov:2011gs}
\begin{equation}
I_{3}(\Delta) =\int \frac{d^{d}xd^{d}yd^{d}z\sqrt{g_{x}}\sqrt{g_{y}}\sqrt{g_{z}}}{[s(x,y)s(y,z)s(z,x)]^{\Delta}}= R^{3(d-\Delta)} \frac{8\pi^{\frac{3(1+d)}{2}}\Gamma(d-
\frac{3\Delta}{2})}{\Gamma(d)\Gamma(\frac{1+d-\Delta}{2})^{3}}\, .
\label{I3}
\end{equation}
For the integrated 4-point function we find
\begin{align}
S_{4}&= 24N(N-1)C_{\psi}^{8} \int \prod_{i=1}^{4}dx_{i}\sqrt{g_{x_{i}}}\bigg(\frac{2 (N^2-3 N+4)}{(s_{13} s_{14} s_{23} s_{24})^{2 d-2}}+\frac{2N}{s_{12}^{2 d-4}  s_{34}^{2 d-4}s_{13}^{d} s_{14}^{d} s_{23}^{d} s_{24}^{d}}\notag\\
&+\frac{2 s_{13}^4 s_{24}^4 s_{12}^d s_{34}^d+s_{12}^{d+4} s_{34}^{d+4}+4 s_{12}^2 s_{34}^2 s_{13}^{d+2} s_{24}^{d+2}-4 s_{12}^4 s_{34}^4 s_{14}^d s_{23}^d-4 s_{14}^2 s_{23}^2 s_{12}^{d+2} s_{34}^{d+2}}{(s_{12} s_{13}^{2 } s_{14}^{2 } s_{23}^{2 } s_{24}^{2 } s_{34})^{d} }\notag\\
&-4(N-1) \frac{s_{12}^{2-2 d} s_{13}^{2 d} s_{34}^{2-2 d} s_{24}^{2 d}+s_{13}^2 s_{24}^2+s_{14}^2 s_{23}^2-s_{12}^2 s_{34}^2}{s_{13}^{3 d-2} s_{14}^{d} s_{23}^{d} s_{24}^{3 d-2}}\bigg),
\end{align}
where $s_{mn}\equiv s(x_{m},x_{n})$ is a chordal distance on a sphere.  
Using the methods described in  \cite{Fei:2015oha, Giombi:2015haa}, we find 
\begin{align}
S_{4}=&C_{\psi}^{8}(2 R)^{4 (2-d)}\frac{2^{1-d} \pi ^{\frac{d+1}{2}} }{\Gamma (\frac{d+1}{2})}\frac{e^{\frac{3}{2} \gamma  (2-d)}}{\left(\pi ^{d/2}\right)^3} 192 N (N-1) \bigg((N-2)^2 \Big(-\frac{4}{\epsilon ^2}+\frac{10}{\epsilon }-\big(25+\frac{7 \pi ^2}{12}\big)\Big)\notag\\
&-2 (N-2) \Big(\frac{3}{5 \epsilon }-\frac{11}{5}\Big)+1+\mathcal{O}(\eps)\bigg)\,.
\end{align}
After expressing the bare coupling $g_{0}$ in terms of the renormalized one
\begin{align}
g_{0}= \mu^{-\epsilon}\bigg(g-\frac{N-2}{2\pi} \frac{g^{2}}{\epsilon}+ \left(\frac{(N-2)^2}{4 \pi ^2 \epsilon ^2}-\frac{N-2}{8 \pi ^2 \epsilon }\right)g^3+ ...\bigg)\,,
\end{align}
we find a surviving pole in $F$ of order $g^4/\eps$. This pole can be cancelled provided we renormalize the Euler density parameter as 
\begin{align}
& b_{0} = \mu^{\epsilon}\bigg(b -\frac{N}{48\pi \epsilon}+\frac{ N (N-1)(N-2) }{30 (4 \pi )^{5}  }\frac{g^{4}}{\epsilon}\bigg)\,,\notag\\
&\beta_{b}= -\epsilon b +\frac{N}{48\pi}-\frac{ N (N-1) (N-2)}{6 (4 \pi )^{5} } g^{4}\,,
\label{betaphi4}
\end{align}
where the coupling independent term is due to the trace anomaly of the free fermion field, and we used $\int d^d x\sqrt{g}{\cal R} = 8\pi+\mathcal{O}(\eps)$.  
In order to obtain the correct expression for $F$ at the UV fixed point in $d=2+\epsilon$, we should now set $g=g^*$ and 
$b=b^*=\frac{N}{48\pi \epsilon}-\frac{ N (N-1)(N-2) }{6 (4 \pi )^{5}  }\frac{g_*^{4}}{\epsilon}$. The contribution of the Euler term 
$\delta F_b = b_0 \int d^d x \sqrt{g}{\cal R}$ at the fixed point (\ref{GNcritcoup}) is then
\begin{equation}
\delta F_b= -\frac{N (N-1) }{48 (N-2)^3} \epsilon ^3 +\mathcal{O}(\epsilon^{4})\,,
\end{equation}
and putting this together with the contributions of $S_2$ and $S_3$,\footnote{Note that even though we had to compute $S_4$ to fix the 
renormalization of $b$, we cannot obtain $F$ to order $\eps^4$. That would require fixing $\beta_b$ to order $g^5$.} we find
\begin{align}
F= N F_f+
\frac{N (N-1)  }{24 (N-2)^2}\epsilon ^2 -\frac{N (N-1) (N-3)}{16 (N-2)^3} \epsilon ^3+\mathcal{O}(\epsilon^{4})\,,
\end{align}
or, in terms of $\tilde{F}= -\sin(\pi d/2)F$:
\begin{align}
\tilde{F}=N\tilde{F}_f+
\frac{ N (N-1) \pi\epsilon ^3}{48 (N-2)^2}-\frac{ N (N-1) ( N-3) \pi\epsilon ^4}{32 (N-2)^3}+\mathcal{O}(\epsilon^{5})\,.
\label{tFGN}
\end{align}
This result agrees with the one obtained in \cite{Fei:2015oha} using a conformal perturbation theory approach; this provides a non-trivial check 
on the procedure used here which involves the curvature terms.

\subsection{Large $N$ expansions}

\label{GNLNsec}

In this section we test the $4-\epsilon$ and $2+\epsilon$ expansions by comparing them with the known $1/N$ expansions
\cite{Gracey:1990wi,Gracey:1992cp,  Gracey:1993kb}. The general form of
the large $N$ expansions of scaling dimensions in the GN model is\footnote{The $1/N^3$ term in $\Delta_{\psi}$ may be found in \cite{Gracey:1993kc, Vasiliev:1992wr}.} 
\begin{align}
&\Delta_{\psi} = \frac{d-1}{2}  + \frac 1 N \gamma_{\psi, 1} + \frac {1} {N^2} \gamma_{\psi, 2} + \mathcal{O}(N^{-3})\, , \notag \\ 
&\Delta_{\sigma} = 1+ \frac 1 N \gamma_{\sigma, 1} + \frac {1} {N^2} \gamma_{\sigma, 2} + \mathcal{O}(N^{-3})\, , \notag \\
&\Delta_{\sigma^2} = 2+ \frac 1 N \gamma_{\sigma^2, 1} + \frac {1} {N^2} \gamma_{\sigma^2, 2} + \mathcal{O}(N^{-3})\,,
\label{GNdimlargeN}
\end{align}
where 
\begin{align}
\gamma_{\psi, 1} &=
-\frac{\Gamma(d-1)}{\Gamma(\frac{d}{2}-1)\Gamma(1-\frac{d}{2})\Gamma(\frac{d}{2}+1)\Gamma(\frac{d}{2})} 
\,, \quad \
\gamma_{\psi, 2} =4 \gamma_{\psi, 1}^2 \left(\frac{(d-1) \Psi (d)}{d-2}+\frac{4 d^2-6 d+2}{(d-2)^2 d}\right)\,, \label{GNpsilargeN}\\
\gamma_{\sigma, 1} &=-4 \frac{d-1}{d-2} \gamma_{\psi, 1}\,, \notag\\
\gamma_{\sigma, 2}&= -\gamma_{\psi, 1}^2 \left(\frac{6 d^2 \Theta (d)}{d-2}+\frac{16 (d-1)^2 \Psi (d)}{(d-2)^2}-\frac{4 (d-1) \left(d^4-2 d^3-12 d^2+20 d-8\right)}{(d-2)^3 d}\right) \label{GNsigmalargeN} 
\end{align}
and 
\begin{align}
\gamma_{\sigma^2,1} &=4(d-1)\gamma_{\psi, 1} \,, \notag\\
\gamma_{\sigma^2,2} &=-\frac{16 d \gamma_{\psi, 1}^2 }{d-2}\bigg(\Big(\frac{1}{d-4}-\frac{d^2}{2}+\frac{4}{(d-4)^2}-\frac{6}{d-2}-\frac{1}{d}-\frac{3}{2}\Big) \Psi (d)+\frac{4}{ (d-4)^2 \gamma_{\psi, 1}}\notag\\
&+\frac{3 d \Theta (d)}{8} \Big(9-d+\frac{12}{d-4}\Big) -\frac{(d-3) d }{d-4}(\Phi (d)+\Psi (d)^2)-\frac{5 d}{2}+\frac{1}{2 (d-4)}-\frac{2}{(d-4)^2}\notag\\
&-\frac{7}{d-2}-\frac{4}{(d-2)^2}+\frac{5}{2 d}-3+\frac{d^2}{2}-\frac{1}{d^2}\bigg)\,.\label{GNsigma2largeN}
\end{align}
In these equations we defined
\begin{align}
&\Psi(d)= \psi(d-1)-\psi(1)+\psi \left (2-\frac {d} {2}\right )-\psi \left (\frac {d} {2}\right )\ , \notag\\
&\Theta(d)=\psi'\left (\frac {d} {2} \right)-\psi'(1)\ , \notag\\
&\Phi(d)= \psi'(d-1)-\psi' \left (2-\frac {d} {2}\right )-\psi' \left (\frac {d} {2} \right )+\psi'(1)
\ ,\end{align}
where $\psi(x)=\Gamma'(x)/\Gamma(x)$ denotes the digamma function. 

In $d=3$, the above large $N$ expansion of the scaling dimensions read
\begin{equation}
\begin{aligned}
\Delta_{\psi}&=1+ \frac{4}{3\pi^2 N} + \frac{896}{27\pi^4 N^2} = 1 +\frac{0.1351}{N}+ \frac{0.3407}{N^2}\,, \\
\Delta_{\sigma}&=1-\frac{32}{3\pi^2 N}+\frac{9728-864\pi^2}{27\pi^4 N^2}= 1-\frac{1.0808}{N}+\frac{0.4565}{N^2}\,, \\
\Delta_{\sigma^2}&=2+\frac{32}{3\pi^2 N}-\frac{64(632+27\pi^2)}{27\pi^4 N^2}=2+\frac{1.0808}{N}-\frac{21.864}{N^2}\,.
\label{lN-3d}
\end{aligned}
\end{equation}

For $\tilde{F}$, the $\mathcal{O}(N^0)$ result can be obtained from the general formula \cite{Diaz:2007an,Giombi:2014xxa} for the change in $F$ 
under a ``double trace" deformation $\delta S = g \int O_{\Delta}^2$, where $O_{\Delta}$ is a scalar primary operator of dimension $\Delta$. In terms 
of $\tilde F=-\sin(\pi d/2)F$, the result is
\begin{equation}
\label{tFLN}
\delta \tilde F = 
\frac{1}{\Gamma\left(d+1\right)}\int_0^{\Delta-\frac{d}{2}} du\,u \sin(\pi u) \Gamma\left(\frac{d}{2}+u\right)\Gamma\left(\frac{d}{2}-u\right)\,.
\end{equation}
In the present case $O_{\Delta} = \bar\psi\psi$, and so $\Delta = d-1$. Therefore, 
\begin{equation}
\label{tFLNGN}
\tilde F = N\tilde F_f + 
\frac{1}{\Gamma\left(d+1\right)}\int_0^{\frac{d}{2}-1} du\,u \sin(\pi u) \Gamma\left(\frac{d}{2}+u\right)\Gamma\left(\frac{d}{2}-u\right) + \mathcal {O} (1/N)\ .
\end{equation}
The $4-\eps$ expansion of this agrees with the large $N$ expansion of (\ref{tFGNYall}), i.e.
\begin{align}
\tilde{F}
=N \tilde F_f + \tilde F_s -\left(\frac{\pi  \epsilon ^2}{96}+\frac{\pi  \epsilon ^3}{192}\right)+\left(\frac{\pi  \epsilon ^2}{16}-\frac{ \pi  \epsilon ^3}{32}\right)\frac{1}{N}-\left(\frac{3 \pi  \epsilon ^2}{8}-\frac{17 \pi  \epsilon ^3}{32}\right)\frac{1}{N^2} +\mathcal{O}(1/N^{3}),
\label{tFGNY}
\end{align}
Similarly, the $2+\eps$ expansion of (\ref{tFLNGN}) agrees with the large $N$ expansion of (\ref{tFGN}), i.e.
\begin{align}
\tilde{F}= N\tilde{F}_f+ \left(\frac{\pi  \epsilon ^3}{48}-\frac{\pi  \epsilon ^4}{32}\right)+\frac{1}{N}\left(\frac{\pi  \epsilon ^3}{16 }-\frac{\pi  \epsilon ^4}{16 }\right)+\frac{1}{N^{2}}\left(\frac{\pi  \epsilon ^3}{6}-\frac{3 \pi  \epsilon ^4}{32}\right)+\mathcal{O}(1/N^{3})\,.
\label{tFGNlargeN}
\end{align}

For $C_T$, the relative $\mathcal{O}(\frac{1}{N})$ correction to the answer for free feermions is given in \cite{Diab:2016spb}:
\begin{align}
& C_{T} = N C_{T,f} \Big(1+\frac{C_{T1}}{N}+\mathcal{O}(1/N^{2})\Big)\ , \notag \\
& C_{T1}= - 4 \gamma_{\psi,1} \left(\frac{\Psi(d)}{d+2}+\frac{d-2}{(d-1)d(d+2)} \right)\,.
\end{align}
Its expansion in $d=4-\eps$ can be seen to match (\ref{CT-GNY}), in particular reproducing the extra propagating scalar present in the GNY description. 

\subsection{Pad\'e approximants}
\label{PadeGN}

To obtain estimates for the CFT observables in $d=3$, we will use ``two-sided" Pad\'e approximants that combine information from the $4-\epsilon$ and 
$2+\epsilon$ expansions. Namely, we consider the rational approximant 
${\rm Pade}_{[m,n]}=\frac{\sum_{i=0}^m a_i d^i}{1+\sum_{j=1}^n b_j d^j}$, 
where $2<d<4$ is the spacetime dimension, and we fix the coefficients $a_i, b_j$ so that its Taylor expansion near $d=4$ and $d=2$ 
agrees with the available perturbative results. Clearly, the ``degree" $n+m$ of the approximant is 
bound by how many terms in the $\epsilon$-expansion are known. Such approximants may be derived for any finite $N$, and 
it is useful to compare their large $N$ behavior as a function of $d$ to the $1/N$-expansion results listed in the previous section. When 
several approximants ${\rm Pade}_{[m,n]}$ are possible for the same quantity, we use such comparisons to large $N$ results to choose the one 
which appears to work best.   

\begin{figure}[h!]
\centering
\includegraphics[width=10cm]{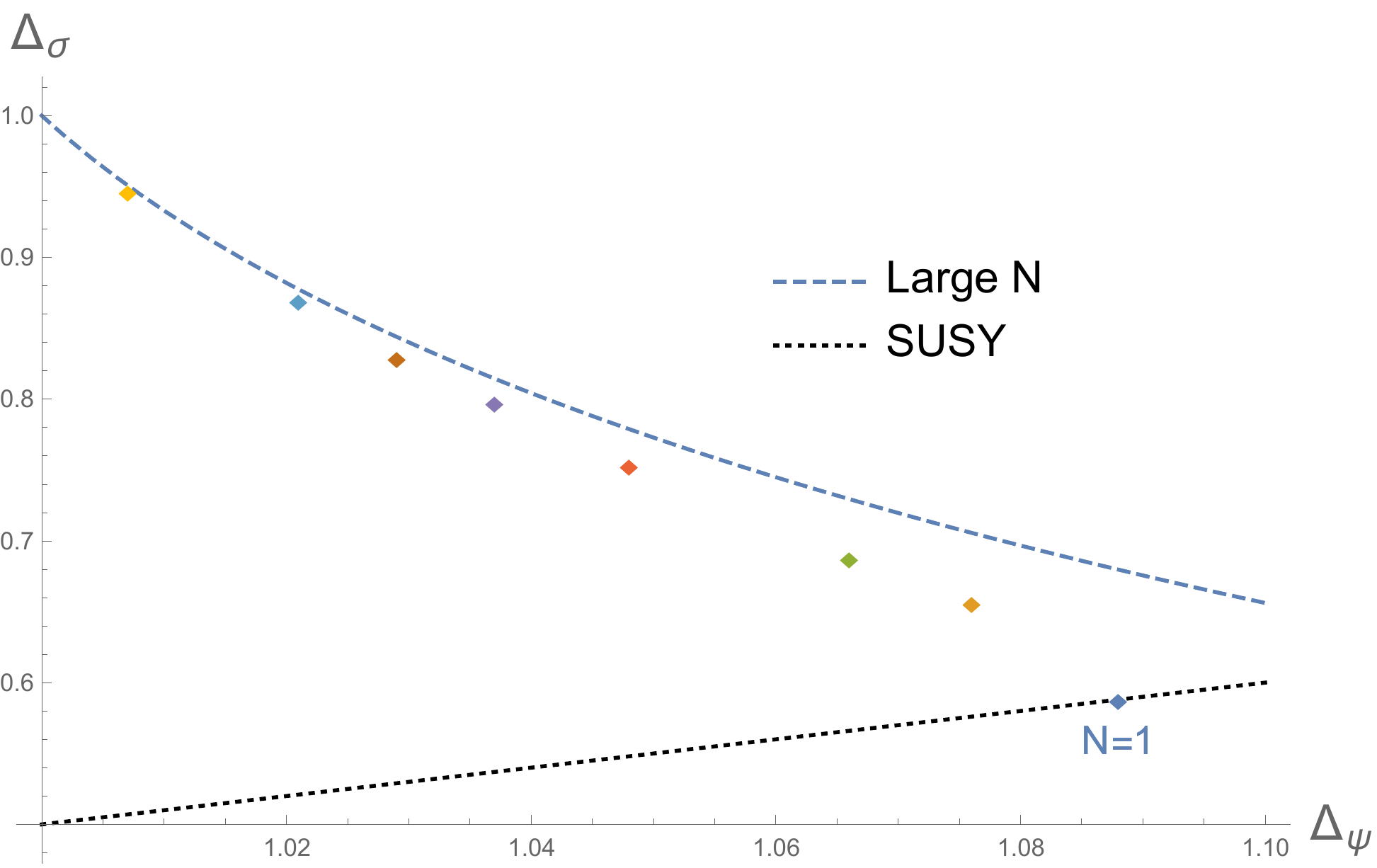}
\caption{Pad\'e estimates in $d=3$ of $\Delta_{\sigma}$ versus $\Delta_{\psi}$ for $N=1,2,3,4,5,6,8,20$, compared to the large $N$ results (\ref{lN-3d}). The $N=1$ value corresponds to the SUSY fixed point discussed in Section \ref{emergentSUSY}. The black dotted line is the SUSY relation $\Delta_{\sigma}=\Delta_{\psi}-1/2$.}
\label{PadevsLN}
\end{figure}

For $\Delta_{\psi}$, $\Delta_{\sigma}$ and $\Delta_{\sigma^2}$ we know the $4-\epsilon$ expansion to order $\eps^2$, and the $2+\eps$ expansion to order $\eps^3$. 
This allows to use Pad\'e approximants with $m+n=6$. For $\Delta_{\psi}$ and $\Delta_{\sigma}$, we find that ${\rm Pade}_{[4,2]}$ 
has no poles in $2<d<4$ and for large $N$ is in good agreement with the results (\ref{GNpsilargeN}) and (\ref{GNsigmalargeN}). For $\Delta_{\sigma^2}$, we perform the Pad\'e on the critical exponent $\nu^{-1} = d-\Delta_{\sigma^2}$, and then translate to $\Delta_{\sigma^2}$ at the end. In this case, we find that ${\rm Pade}_{[1,5]}$ is the only 
approximant with no poles, and it matches well to the large $N$ result. The resulting estimates in $d=3$ for these scaling dimensions 
are given in Table \ref{3dGN-table}. We note that the scaling dimensions for $N=8$ are in good agreement with the Monte Carlo results from \cite{Karkkainen:1993ef}. 
In figure \ref{PadevsLN}, we plot our 3d estimates for $\Delta_{\psi}$ and $\Delta_{\sigma}$, compared to the large $N$ 
curve obtained from (\ref{lN-3d}) by eliminating $N$ to express $\Delta_{\sigma}$ as a function of $\Delta_{\psi}$. The $N=1$ values, which correspond to 
the ${\cal N}=1$ SUSY fixed point, are obtained in Section \ref{emergentSUSY}. 

\begin{table}[h]
\centering
\begin{tabular}{cccccccc}
\hline
\multicolumn{1}{|c|}{$N$}         & \multicolumn{1}{c|}{3} & \multicolumn{1}{c|}{4} & \multicolumn{1}{c|}{5} & \multicolumn{1}{c|}{6}& \multicolumn{1}{c|}{8} & \multicolumn{1}{c|}{20}& \multicolumn{1}{c|}{100} \\ \hline
\multicolumn{1}{|c|}{$\Delta_\psi$ (Pade$_{[4,2]}$)}      & \multicolumn{1}{c|}{1.066} & \multicolumn{1}{c|}{1.048} & \multicolumn{1}{c|}{1.037} & \multicolumn{1}{c|}{1.029} & \multicolumn{1}{c|}{1.021}& \multicolumn{1}{c|}{1.007} & \multicolumn{1}{c|}{1.0013}  \\ \hline
\multicolumn{1}{|c|}{$\Delta_\sigma$ (Pade$_{[4,2]}$)} & \multicolumn{1}{c|}{0.688} & \multicolumn{1}{c|}{0.753} & \multicolumn{1}{c|}{0.798} & \multicolumn{1}{c|}{0.829} & \multicolumn{1}{c|}{0.87} & \multicolumn{1}{c|}{0.946} & \multicolumn{1}{c|}{0.989} \\ \hline
\multicolumn{1}{|c|}{$\Delta_{\sigma^2}$ (Pade$_{[1,5]}$)}      & \multicolumn{1}{c|}{2.285} & \multicolumn{1}{c|}{2.148} & \multicolumn{1}{c|}{2.099} & \multicolumn{1}{c|}{2.075} & \multicolumn{1}{c|}{2.052}& \multicolumn{1}{c|}{2.025}& \multicolumn{1}{c|}{2.008}  \\ \hline
\multicolumn{1}{|c|}{$F/(N F_f)$ (Pade$_{[4,4]}$)} & \multicolumn{1}{c|}{1.091} & \multicolumn{1}{c|}{1.060} & \multicolumn{1}{c|}{1.044} & \multicolumn{1}{c|}{1.034} & \multicolumn{1}{c|}{1.024} & \multicolumn{1}{c|}{1.008}  & \multicolumn{1}{c|}{1.0014} \\ \hline
\multicolumn{1}{l}{}            & \multicolumn{1}{l}{}   & \multicolumn{1}{l}{}   & \multicolumn{1}{l}{}   & \multicolumn{1}{l}{}   & \multicolumn{1}{l}{}  & \multicolumn{1}{l}{} & \multicolumn{1}{l}{}
\end{tabular}
\caption{Estimates of scaling dimensions and sphere free energy at the $d=3$ interacting fixed point of the GN model.}
\label{3dGN-table}
\end{table}

\begin{figure}[h!]
\centering
\includegraphics[width=8cm]{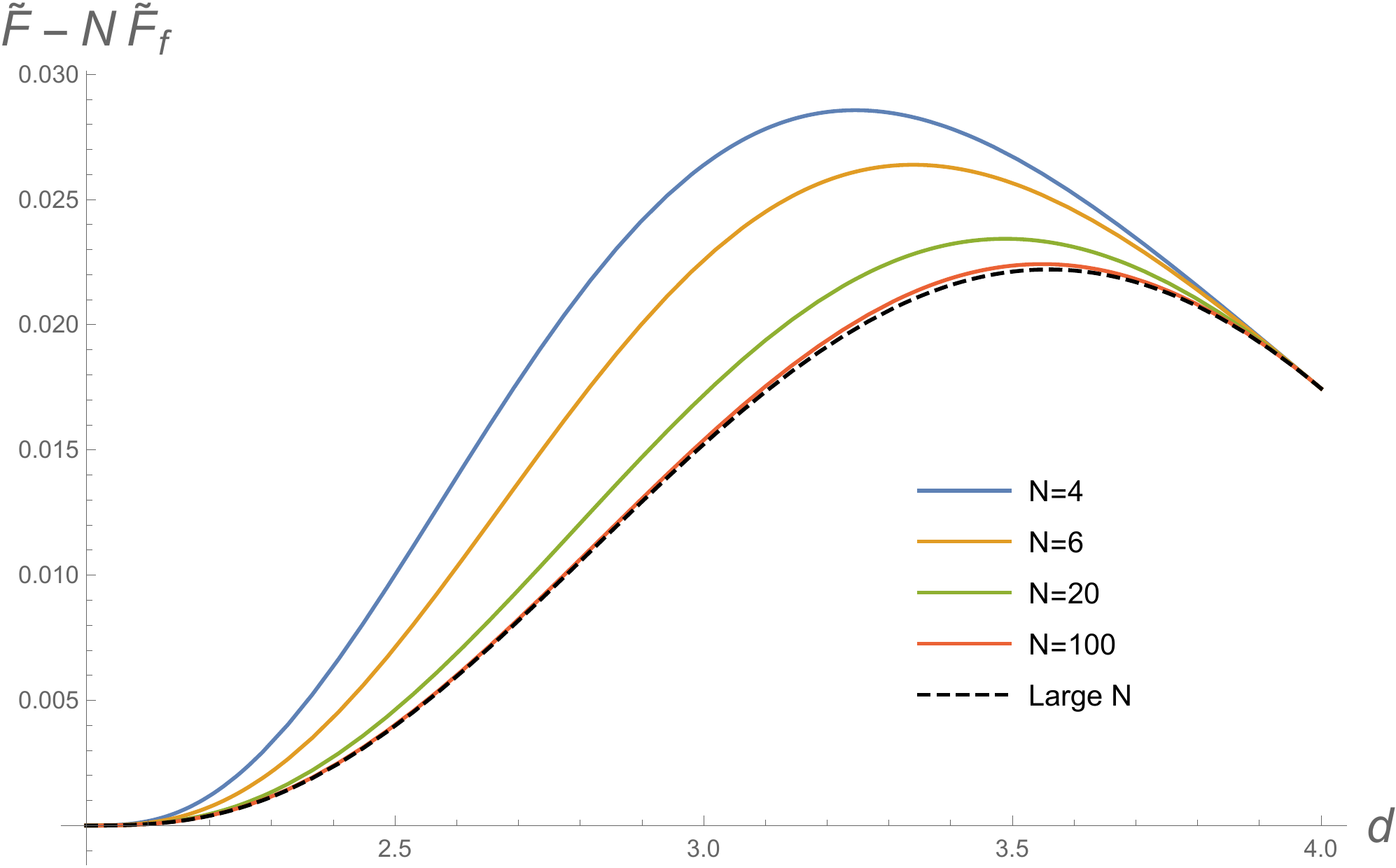}
\caption{Pad\'e estimates of $\tilde{F}-N\tilde F_f$ in $2<d<4$ compared to the large $N$ result (\ref{tFLN}).}
\label{F-PadevsLN}
\end{figure}

For the sphere free energy of the interacting CFT, $\tilde F$, we find it convenient to perform the Pad\'e approximation on the quantity 
\begin{equation}
f(d) = \tilde F - N \tilde F_{f} \,,
\end{equation}
which is essentially the interacting part of the free energy in the GN description, but it includes the contribution of a free scalar 
from the GNY point of view. Using the results (\ref{tFGNYall}) and (\ref{tFGN}), we can use Pad\'e approximants with $n+m=8$, 
and we find that ${\rm Pade}_{[4,4]}$ has no poles and agrees well with the large $N$ result (\ref{tFLN}). The resulting $d=3$ estimates for 
$\tilde F$, normalized by the free fermion contribution $NF_f$, are given in Table \ref{3dGN-table} for a few values of $N$. In Figure \ref{F-PadevsLN}, we also plot 
the result of the constrained Pad\'e approximants for $\tilde F-N \tilde F_f$ as a function of $2<d<4$ for a few values of $N$, showing that they approach well the analytical large $N$ formula (\ref{tFLN}).

We can use our estimates to make some tests of the $F$-theorem. In the GNY description, we can flow to the critical theory from the free UV fixed point 
of $N$ fermions plus a scalar, while in the GN description one can flow to the critical theory to the free fermions. Then, the $F$-theorem 
implies the inequalities 
\begin{equation}
N F_f + F_s > F > N F_f \,.
\label{Fth1}
\end{equation}
We verified that our estimates satisfy these inequalities for all values of $N$. As an example, for $N=4$ we get $F_{\textrm{GN}}/(N F_f+F_s) \approx 0.93$ 
and  $F_{\textrm{GN}}/(N F_f) \approx 1.06$. In the GNY description, we also see that we can flow to the critical GN point 
from the decoupled product of the Ising CFT and $N$ free fermions. This implies 
\begin{equation}
 N F_f + F_{\rm Ising} > F \, .
\label{Fth2}
\end{equation}
Using the estimates for $F_{\rm Ising}$ derived in \cite{Fei:2015oha}, we have checked that this inequality indeed holds. Using the Pad\'e 
approximants as a function of $2<d<4$, we can also verify that both (\ref{Fth1}) and (\ref{Fth2}) are satisfied, in terms of $\tilde F$, 
in the whole range of $d$. This is in agreement with the ``generalized $F$-theorem" \cite{Fei:2015oha}. 

Finally, we discuss $N=2$, which is a special case where the $\beta$-function in $d=2$ vanishes exactly; therefore, the theory has a line of fixed points.
For $N=2$ we cannot apply the strategy described above because the $2+\eps$ expansions (\ref{GNdim}) become singular. 
Directly in $d=2$, the GN model is equivalent to the
Thirring model for a single 2-component Dirac fermion which can be solved via bosonization and has a line of fixed points; 
the dimensions of $\psi$ and $\sigma=\bar \psi\psi$ depend on the interaction strength. Therefore, for these operators we can only perform  the ``one-sided" ${\rm Pade}_{[1,1]}$ on the $4-\eps$ expansions.
In $d=2$ it is known that $\Delta_{\sigma^2}=2$ (the operator is exactly marginal) and $c=1$; we impose these boundary conditions on the Pad\' e approximants. 
Then the results in $d=3$ are 
\begin{equation}
\Delta_\psi^{N=2} \approx 1.076\,, \qquad 
\Delta_\sigma^{N=2} \approx 0.656\,, \qquad 
\Delta_{\sigma^2}^{N=2} \approx 1.75 \,,
\label{ntwoest}
\end{equation}
and for the sphere free energy
\begin{equation}
F^{N=2} \approx 0.254 \approx 1.16(2F_f)  \approx 0.9 (2F_f+F_s)\,,
\end{equation}
in agreement with the $F$-theorem.

\section{The Nambu-Jona-Lasinio-Yukawa Model}
\label{NJLYsection}

 Using the results of \cite{Machacek:1983tz, Machacek:1983fi, Machacek:1984zw,Jack:1990eb,Pernici:1999nw}, 
we have found the following $\beta$-functions for the NJLY model (\ref{NJLYlagrange}) up to two loops:
\begin{align}
\beta_1 &= -\frac{\eps}{2}g_1 + \frac{1}{(4\pi)^2}\left (\frac{N}{2}+2\right )g_1^3+ \frac{1}{(4\pi)^4}\left (-\frac{8}{3} g_1^3 g_2 + \frac{1}{9} g_1 g_2^2 +
(7-3N) g_1^5 \right )\ , \notag \\
\beta_2&= -\eps g_2+\frac{1}{(4\pi)^2}\Big(\frac{10}{3}g_2^2+2N g_2 g_1^2-12N g_1^4\Big) 
- \frac{1}{(4\pi)^4}\left (\frac{20}{3} g_2^3  - 96 N g_1^6 - 2 N g_2 g_1^4 +\frac{10}{3} N g_1^2 g_2^2 \right )
\ .\label{NJLYbeta}
\end{align}
The one-loop terms above agree with the $N_b=1$ case of the results in \cite{Roy:2012wz}.  
Solving (\ref{NJLYbeta}), we find the following fixed point:
\begin{align}
&\frac {(g_{1}^{*})^{2}} {(4\pi)^{2}}  = \frac{\eps}{N+4} + \frac{-N^2+448 N-1096+ (N+76) \sqrt{N^2+152N+16}}{100 (N+4)^3} \eps^2 + \mathcal{O}(\eps^3)\ , \notag \\
& \frac{g_{2}^{*}} { (4\pi)^{2}}= \frac{3(\sqrt{N^2+152N+16}-N+4)}{20(N+4)}\eps +\frac{9}{500(N+4)^3\sqrt{N^2+152N+16}} \notag \\
& ~~~~~~~~~~~~~\times \Big(3N^4+114N^3+764N^2-26192N+1280\notag\\
&~~~~~~~~~~~~~~~~~~~-(3N^3-114N^2-1932N-320)\sqrt{N^2+152N+16}\Big)\eps^2 +\mathcal{O}(\eps^3)\, . 
\label{NJLYcoupl}
\end{align}
To get a positive solution for $g_2^*$ we have picked the $+$ sign for the square root. The other choice of sign gives another fixed point which is 
presumably unstable. In addition, there is clearly a fixed point with $g_1^*=0$ and $g_2^*$ corresponding to the $O(2)$ Wilson-Fisher fixed point. The 
solution (\ref{NJLYcoupl}) yields an IR stable fixed point for all values of $N$.

The scaling dimensions of the fields are found to be
\begin{align}
\Delta_\psi&= \frac{3-\eps}{2} +\frac{1}{(4\pi)^2}g_1^2 
-\frac{1}{4 (4\pi)^4}  (3N +2) g_1^4
\ ,\notag \\
\Delta_\phi &= 1-\frac{\eps}{2} + \frac{1}{2 (4\pi)^2} Ng_1^2+ \frac{1}{(4\pi)^4}\left (\frac{1}{9} g_2^2 
-\frac{3}{2} N g_1^4 \right )\, .
\label{NJLYdim}
\end{align}
At the fixed point (\ref{NJLYcoupl}) these give:
\begin{align}
\Delta_{\psi} &=\frac{3}{2} - \frac{N+2}{2(N+4)} \eps + \frac{-76N^2 + 98N -1296 + (N+76)\sqrt{N^2+152N+16}}{100(N+4)^3}\eps^2 +\mathcal{O}(\eps^3)\ , \notag \\
\Delta_{\phi} &=1- \frac{2}{N+4} \eps + \frac{56N^2-498N+16+(19N+4)\sqrt{N^2+152N+16}}{50(N+4)^3}\eps^2 +\mathcal{O}(\eps^3)\ .
\label{NJLYdimcrit}
\end{align}

The NJLY has two types of operators quadratic in the scalar fields: the $U(1)$ invariant operator $\phi \bar \phi$, and the charged
operators $\phi^2$ and $\bar \phi^2$.
The one-loop scaling dimension of $\phi\bar \phi$ 
was determined in \cite{Roy:2012wz}.
Using \cite{Pernici:1999nw}, we find up to two loops 
\begin{align}
&\Delta_{\phi \bar \phi}  = d-2 +\frac{4}{3(4\pi)^{2}}g_{2} - \frac{4}{3(4\pi)^{4}}\Big(g_{2}^{2}+ N g_{2}g_{1}^{2}\Big)+2\gamma_{\phi}\, ,\notag \\
& \Delta_{\phi^2 }  = d-2 +\frac{2}{3(4\pi)^{2}}g_{2}- \frac{2}{3(4\pi)^{4}}\Big(\frac {4} {3} g_{2}^{2}+ N g_{2}g_{1}^{2}- 6 N g_{1}^{4} \Big )+
2\gamma_{\phi}\ . \label{NJLsigma2}
\end{align}
At the fixed point this gives
\begin{align}
\Delta_{\phi \bar \phi} =&2+ \frac{\sqrt{N^2+152N+16}-N-16}{5(N+4)}\eps + \frac{1}{250 (N+4)^3 \sqrt{N^2+152N+16}} \notag\\
& \times \Big((17 N^3+ 104 N^2 +1252 N+1120) \sqrt{N^2+152N+16} \notag\\
&~~~~~~- 17 N^4- 4646 N^3+ 2304 N^2- 187712 N  + 4480  \Big) \eps^2 +\mathcal{O}(\eps^3)\,
\label{NJLYsigsq}
\end{align}
and
\begin{align}
\Delta_{\phi^2} =&2+ \frac{\sqrt{N^2+152N+16}-N-36}{10(N+4)}\eps + \frac{1}{125 (N+4)^3 \sqrt{N^2+152N+16}} \notag\\
& \times \Big((3 N^3+ 571 N^2 +688 N+240) \sqrt{N^2+152N+16} \notag\\
&~~~~~~ - 3 N^4- 924 N^3+ 7806 N^2- 47688 N  + 960  \Big) \eps^2 +\mathcal{O}(\eps^3)\,.
\label{NJLYsigsqcharge}
\end{align}

The $4-\epsilon$ expansion of $C_T$ in the NJLY model proceeds similarly to that for the GNY model presented in \cite{Diab:2016spb} and reviewed in
section \ref{GNYsection}.
In the NJLY model there are two real scalar fields, and we need to replace $T_\sigma$ by  $T_{\sigma_1}+ T_{\sigma_2}$.
It is not hard to check that each  diagram contributing to the term $\sim g_1^2$ picks up a factor of 2 compared to the GNY model, since each of the internal scalar lines
can be either $\phi_1$ or $\phi_2$.
Thus, we find
\begin{align}
C_T = N C_{T,f}+ 2 C_{T,s}- \frac {5 N (g_1^*)^2}{6 (4 \pi)^2} 
= \frac{d}{S_d^2}\left(\frac{N}{2} + \frac{2}{d-1} - \frac{5N\epsilon}{6(N+4)}\right)\, .
\label{CTNJLY}
\end{align}

\subsection{Free energy on $S^{4-\eps}$}

The calculation of $F$ for the NJLY model follows the same steps as the one for the GNY model discussed earlier. The integrals 
are also nearly identical, 
except for combinatorics factors due to the fact that we have two scalar fields, one of which has $i\gamma^5$ coupling.

The perturbative expansion of the free energy is given by
\begin{align}
F&=N F_f+2 F_s -\frac{1}{2}g_{1,0}^{2} \int d^{d}xd^{d}y \sqrt{g_{x}}\sqrt{g_{y}} \langle \mathcal{O}_1(x) \mathcal{O}_1(y)\rangle_{0}\notag\\
&~~~~~- \frac{1}{2!(4!)^{2}}g_{2,0}^{2}\int d^{d}xd^{d}y \sqrt{g_{x}}\sqrt{g_{y}} \langle \mathcal{O}_2(x) \mathcal{O}_2\rangle_{0} \notag\\
&~~~~~-\frac{1}{4!}g_{1,0}^{4}
\int d^{d}xd^{d}yd^{d}zd^{d}w \sqrt{g_{x}}\sqrt{g_{y}}\sqrt{g_{z}}\sqrt{g_{w}}\langle  \mathcal{O}_1(x) \mathcal{O}_1(y)  \mathcal{O}_1(z) \mathcal{O}_1(w)\rangle_{0} + \delta F_b \,,
\end{align}
where we defined the operators $\mathcal{O}_1 =\bar{\psi_j}(\phi_1+i\gamma_5 \phi_2)\psi^j$ 
and $\mathcal{O}_2 =(\phi_1^2+\phi_2^2)^2$, and $\delta F_b = b_0 \int d^d x E$ is the Euler term. Evaluating the correlation functions above, 
we find
\begin{equation}
F=N F_f + 2 F_s -\frac{1}{2}2g_{1,0}^{2} N C_{\phi}C_{\psi}^{2}  I_{2}\big(\frac{3}{2}d-2\big)
+ \frac{1}{2\cdot (4!)^2}64g_{2,0}^{2}C_{\phi}^{4}I_{2}(2d-4) +\frac{1}{4!}g_{1,0}^{4}I_4 +\delta F_b\,,
\label{NJLYFree}
\end{equation}
where $I_2$ is given in (\ref{I2}) and $I_4$ corresponds to the integrated 4-point function of ${\cal O}_1$, for which we find
\begin{align}
I_4 &= \frac{N(N+4)}{(4\pi)^4 \eps}+ \frac{N\left(19N+72 + 6(N+4)(\gamma+\log(4\pi R^2))\right)}{6(4\pi)^4}+\mathcal{O}(\eps)\,.
\end{align}
Now replacing the bare couplings with the renormalized ones
\begin{align}
&g_{1,0} = \mu^{\frac{\epsilon}{2}}\Big(g_{1}+\frac{N+4}{32\pi^{2}}\frac{g_{1}^{3}}{\epsilon}+\dots\Big), \quad g_{2,0}= \mu^{\epsilon}\big(g_{2}+\dots \big)
\end{align}
we find that all poles cancel as they should. As explained in the GNY and GN calculations, in order 
to obtain the correct expression for $F$ at the IR fixed point in $d=4-\epsilon$ we need to include the effect of the Euler term. 
Using an improved version of the result from \cite{Jack:1990eb, Jack:1984vj}, adapted to the presence of $\gamma_5$ in the vertices, we get:
\begin{align}
&\beta_{b}= \epsilon b -\frac{11N/4+2}{360(4\pi)^2}-\frac{1}{(4\pi)^8}\frac{1}{1296}\left(N g_1^2 (24 g_1^2 g_2 - g_2^2 + 9g_1^4(3N-7))\right)\,.
\label{betabNJL}
\end{align}
From this we can solve for the fixed point value $b_*$ with the couplings given in (\ref{NJLYcoupl}), and we find that the Euler term 
contributes
\begin{align}
\delta F_b =\frac{N(-1096+76\sqrt{N^2+152N+16}+N(448-N+\sqrt{N^2+152N+16}))}{7200(N+4)^3}\eps^2 \,.
\label{FbNJLY}
\end{align}
Putting this together with the integrals in (\ref{NJLYFree}), we finally find in terms of $\tilde F$:
\begin{align}
&\tilde{F}
=N \tilde F_f + 2 \tilde F_s -\frac{N \pi \epsilon^2 }{48 (N+4)} - \frac{1}{14400 (N+4)^3}\Big(149 N^3 + 2372 N^2 + 1312N + 64\notag\\
& ~~~~~~~~~~~~+ \left(N^2+152 N+16\right)\sqrt{N^{2} +152N+16} \Big)\pi \epsilon^3
+\mathcal{O}(\eps^{4})\,. \label{tFNJLY}
\end{align}

\subsection{$2+\epsilon$ expansions}

In this section we consider the theory in $2+\epsilon$ dimensions with 4-fermion interactions which respect the $U(1)$ chiral symmetry.  
We begin with the action in Euclidean space of the form \cite{Bondi:1989nq}
\begin{align}
S_{\textrm{SPV}} = -\int d^{d}x\Big(\bar{\psi}_{i}\slashed\partial  \psi^{i} + \frac{1}{2}g_{S}(\bar{\psi}_{i}\psi^{i})^{2} +\frac{1}{2}g_{P} (\bar{\psi}_{i}\gamma_{5}\psi^{i})^{2} +\frac{1}{2} g_{V} (\bar{\psi}_{i}\gamma_{\mu}\psi^{i})^{2}\Big) \,,
\end{align}
where 
$i=1,\dots 2 N_f$ 
is the number of two-component spinors, and  $\gamma_{0}=\sigma_{1}$, $\gamma_{1}=\sigma_{2}$ and $\gamma_{5}= -i\gamma_{0}\gamma_{1}=\sigma_{3}$. For $g_V=0$ and $g_S=- g_P$ this reduces to the well-known chiral Gross-Neveu model  \cite{Gross:1974jv} in $d=2$. However, as we will see below, for our purposes
it is not consistent to set $g_V=0$ -- the corresponding operator
respects the $U(1)$ chiral symmetry and gets induced.

The one-loop beta-functions and anomalous dimension of the $\psi$ field were found in \cite{Bondi:1989nq} using MS scheme and  read 
\begin{align}
&\beta_{S}=\eps g_{S}-\frac{1}{\pi}\big(\frac{1}{2}(N-2)g_{S}^2 -g_{P} g_{S}-2 g_{V} (g_{S}+g_{P})\big)+\dots\,, \notag\\
&\beta_{P}=\eps g_{P}+\frac{1}{\pi}\big(\frac{1}{2}(N-2)g_{P}^2 -g_{P} g_{S}+2 g_{V} (g_{S}+g_{P})\big)+\dots\,, \label{betasNJLd2} \\
&\beta_{V}=\eps g_{V}+\frac{1}{\pi }g_{P} g_{S}+\dots\,,   \notag
\end{align}
and
\begin{align}
&\Delta_{\psi}=\frac{1}{2}+\frac{1}{2}\eps+\frac{N }{4(2 \pi )^2 }\left(g_{P}^2+g_{S}^2+2 g_{V}^2\right)-\frac{1}{4 (2 \pi )^2}\big(4 g_{V} (g_{S}-g_{P})+(g_{S}+g_{P})^2\big)\, .
\end{align}
We note that at the leading order in the $2+\eps$ expansion the evanescent operators do not appear \cite{Bondi:1989nq, Vasiliev:1996rd, Vasiliev:1996nx}. 
One of the UV fixed points of (\ref{betasNJLd2}) is\footnote{Equations  (\ref{betasNJLd2}) have four different non-trivial fixed points. Two of them correspond to the GN and $SU(2N_{f})$ Thirring models.} 
\begin{align}
g^{*}_{S}= -g^{*}_{P}=\frac{N}{2}g^{*}_{V}=\frac{2 \pi  \epsilon }{N}\,,
\label{NJLfixedpoint}
\end{align}
which corresponds to the $SU(2 N_f)$ Thirring model  \cite{Bondi:1989nq, Bondi:1988fp}. Indeed  using the relation for the $SU(2N_f)$ generators  $(T^{a})^{i}_{j}(T^{a})^{k}_{l}=\frac{1}{2}(\delta^{i}_{l}\delta_{j}^{k}-\frac{2}{N}\delta^{i}_{j}\delta^{k}_{l})$ and the Fierz identity in $d=2$ $(\gamma_{\mu})_{\alpha\beta}(\gamma_{\mu})_{\gamma\delta} =\delta_{\alpha\delta}\delta_{\beta\gamma}-(\gamma_{5})_{\alpha\delta}(\gamma_{5})_{\beta\gamma}$
one finds
\begin{align}
(\bar{\psi}\psi)^{2}- (\bar{\psi}\gamma_{5}\psi)^{2} +\frac{2}{N}(\bar{\psi}\gamma_{\mu}\psi)^{2} = - 2(\bar{\psi}\gamma_{\mu}T^{a}\psi)^{2}\,.
\label{NJLint}
\end{align}
So the action for this model is 
\begin{align}
S_{SU(2N_{f}) \textrm{Thirring}} = -\int d^{d}x \Big(\bar{\psi}_{i}\slashed\partial  \psi^{i} + \frac{1}{2}g\big((\bar{\psi}_{i}\psi^{i})^{2} - (\bar{\psi}_{i}\gamma_{5}\psi^{i})^{2} + \frac{2}{N} (\bar{\psi}_{i}\gamma_{\mu}\psi^{i})^{2}\big)\Big) \,,
\end{align}
with the beta-function and anomalous dimension  
\begin{align}
\beta_{g} = \eps g -\frac{N}{2\pi}g^{2}+\mathcal{O}(g^{3}), \quad &\Delta_{\psi}=\frac{1}{2}+\frac{1}{2}\eps + \frac{N^{2}-4}{8\pi^{2}N}g^{2}+\mathcal{O}(g^{3})\,.
\end{align}
It is plausible that this model is the continuation of the NJLY model (\ref{NJLYlagrange}) to $d=2+\eps$. One finds that  the UV fixed point is $g_{*}= 2\pi \eps/N$, and the critical anomalous dimension reads 
\begin{align}
\Delta_{\psi} = \frac{1}{2}+\frac{1}{2}\eps + \frac{1}{2}\Big(\frac{1}{N}-\frac{4}{N^{3}}\Big)\eps^{2}  +\mathcal{O}(\eps^{3})\,.
\label{NJLDelpsi}
\end{align}
Also one finds that the dimension of the quartic operator is 
\begin{align}
\Delta_{\phi \bar \phi }=2+\eps+\beta'(g_{*}) = 2+\mathcal{O}(\eps^{2}) \,.
\label{twosigsq}
\end{align}

We can also calculate the free energy of this model. To order $g^2$, we have:
\begin{equation}
F=N F_f -\frac{1}{2!}\Big(\frac{g_{0}}{2}\Big)^{2}S_{2}+{\cal O}(g_0^3)\,,
\end{equation}
where $F_f$ is the free fermion contribution and
\begin{align}
S_{2} = \int dx_1 dx_2 \sqrt{g_{x_{1}}} \sqrt{g_{x_{2}}}\langle\mathcal{O}(x_{1}) \mathcal{O}(x_2)\rangle_{0}^{\textrm{conn}}\,,
\end{align}
where, under the Thirring description (\ref{NJLint}), $\mathcal{O}= - 2(\bar{\psi}\gamma_{\mu}T^{a}\psi)^{2}$.
Going through the combinatorics, we find
\begin{align}
&S_{2}=4 (N^2-4)C_{\psi}^{4}I_{2}(2d-2) \, .
 \end{align}
Evaluating this in $d=2+\eps$ and plugging in the fixed 
point value (\ref{NJLfixedpoint}) we finally find
\begin{equation}
\tilde{F} = N \tilde{F}_f + \frac{(N^2-4)\pi \eps^3}{24 N^2}+{\cal O}(\eps^4)\,.
\label{FintNJL}
\end{equation}

\subsection{Large $N$ expansions}

For the NJL model, the $1/N$ expansions of operator dimensions again assume the general form (\ref{GNdimlargeN}), 
where now  \cite{Gracey:1993cq,Gracey:1994ux}
\begin{align}
\gamma_{\psi, 1} &=-\frac{2\Gamma(d-1)}{\Gamma(\frac{d}{2}-1)\Gamma(1-\frac{d}{2})\Gamma(\frac{d}{2}+1)\Gamma(\frac{d}{2})}\ , \quad \gamma_{\psi, 2}  = 2 \gamma_{\psi,1}^{2} \Big(\Psi(d)+\frac{4}{d-2}+\frac{1}{d}\Big)\,, \label{NJLpsilargeN}\\
\gamma_{\phi, 1} &=-2\gamma_{\psi, 1}, \quad  \gamma_{\phi, 2} = -2 \gamma_{\psi,2} +\frac{4d^{2}(d^{2}-5d+7)}{(d-2)^{3}}\gamma_{\psi,1}^{2}  \label{NJLsigmalargeN} 
\end{align}
and 
\begin{align}
\gamma_{\phi \bar \phi, 1} &=4(d-1)\gamma_{\psi, 1} \,, \\
\gamma_{\phi \bar\phi , 2}& = -8 \gamma_{\psi,1}^{2}\bigg(\frac{3 d^2 (3 d-8) \Theta(d) }{4 (d-4) (d-2)}-\frac{(d-3) d^2 \left(\Phi(d) +\Psi ^2(d)\right)}{(d-4) (d-2)}+\frac{4 (d-2) d}{(d-4)^2 \gamma_{\psi,1}}+\frac{4}{d-2}+\frac{1}{d} \notag\\
&+\frac{(d-4) d^2}{d-2}-\frac{(d-3)^2 d^2}{(d-4)^2 (d-2)}+\Psi(d)\Big(\frac{(d-3) d^2 \left(d^2-8 d+20\right)}{(d-4)^2 (d-2)^2}-d^{2}+1\Big)\bigg)\ .
\label{NJLsigma2largeN}
\end{align}
There is a considerable similarity between these results and the corresponding results for the GN model. For example, $\gamma_{\psi, 1}$ in the NJL model is 2 times that
in the GN model. This is because the NJL lagrangian (\ref{NJLlag}) contains two separate double-trace operators, and hence there are two auxiliary fields, $\phi_1$ and $\phi_2$, that
couple to the fermions. A similar factor of 2 appears in various other quantities.

When (\ref{NJLsigmalargeN}) and (\ref{NJLsigma2largeN}) are expanded in $d=4-\eps$ and $d=2+\eps$, they are consistent with our $\eps$-expansion results.\footnote{
The form of (\ref{twosigsq}) agrees with the large $N$ result (\ref{NJLsigma2largeN}), but we haven't  compared the coefficient of $\eps^2$.}
Setting $d=3$ in these equations gives \cite{Gracey:1993cq,Gracey:1994ux}
\begin{equation}
\begin{aligned}
\Delta_{\psi}&=1+ \frac{8}{3\pi^2 N} + \frac{1280}{27\pi^4 N^2} = 1 +\frac{0.2702}{N}+ \frac{0.4867}{N^2}\,, \\
\Delta_{\phi}&=1-\frac{16}{3\pi^2 N}+\frac{4352}{27\pi^4 N^2}= 1-\frac{0.5404}{N}+\frac{1.6547}{N^2} \,, \\
\Delta_{\phi \bar\phi}&=2+\frac{64}{3\pi^2 N}-\frac{128(364+27\pi^2)}{27\pi^4 N^2}=2+\frac{2.1615}{N}-\frac{30.684}{N^2} \,.
\end{aligned}
\label{lN-3d-NJL}
\end{equation}

Turning to the sphere free energy, we can obtain the $\mathcal{O}(N^0)$ result for $\tilde{F}$ by simply doubling the $\delta \tilde F$ from eq. (3.8) of \cite{Giombi:2014xxa}.
Therefore, we have
\begin{equation}
\label{tFLNNJL}
\tilde F = N\tilde F_f + 
\frac{2}{\Gamma\left(d+1\right)}\int_0^{\frac{d}{2}-1} du\,u \sin(\pi u) \Gamma\left(\frac{d}{2}+u\right)\Gamma\left(\frac{d}{2}-u\right) + \mathcal {O} (1/N)\ .
\end{equation}
The $4-\eps$ expansion of this agrees with the large $N$ expansion of (\ref{tFNJLY}), i.e.
\begin{align}
\tilde{F}
= N \tilde F_f + 2 \tilde F_s -\left(\frac{\pi  \epsilon ^2}{48}+\frac{\pi  \epsilon ^3}{96}\right)+\left(\frac{\pi  \epsilon ^2}{12}-
\frac{ \pi  \epsilon ^3}{18}\right)\frac{1}{N}-\left(\frac{ \pi  \epsilon ^2}{3}-\frac{17 \pi  \epsilon ^3}{36}\right)\frac{1}{N^2} +\mathcal{O}(1/N^{3}),
\label{tFNJLYlarge}
\end{align}
and its $2+\eps$ expansion agrees with the large $N$ expansion of (\ref{FintNJL}), which yields
\begin{equation}
\tilde F = N \tilde F_f +\frac{\pi \eps^3}{24}-\frac{\pi\eps^3}{6N^2}+{\cal O}(1/N^4)\,.
\end{equation}

For $C_T$, the presense of the extra scalar field compared to the GN case \cite{Diab:2016spb} again poses no difficulty. 
After some simple excercise commuting $\gamma^5$, we conclude that all the diagrams in \cite{Giombi:2014xxa} contributing to $C_T$ in the GN model 
should receive a factor of 2 due to the presence of two scalar fields. Hence, we find for the NJL model 
\begin{align}
& C_{T} = N C_{T,f} \Big(1+\frac{C_{T1}}{N}+\mathcal{O}(1/N^{2})\Big)\ , \notag \\
& C_{T1}= - 4 \gamma_{\psi,1} \left(\frac{\Psi(d)}{d+2}+\frac{d-2}{(d-1)d(d+2)} \right)\,,
\end{align}\
i.e. the correction $C_{T1}$ is just twice the corresponding term in the GN model (note that $\gamma_{\psi,1}$ 
in the NJL model is twice that of the GN model). This result is, in particular, consistent with the fact that 
in the limit $d\rightarrow 4$ we expect $C_{T1}$ to reproduce the contribution of two free scalar fields. Its $4-\eps$ expansion 
can be also seen to agree with (\ref{CTNJLY}). 

\subsection{Pad\' e approximants}
\label{PadeNJL}

\begin{figure}[h!]
\centering
\includegraphics[width=10cm]{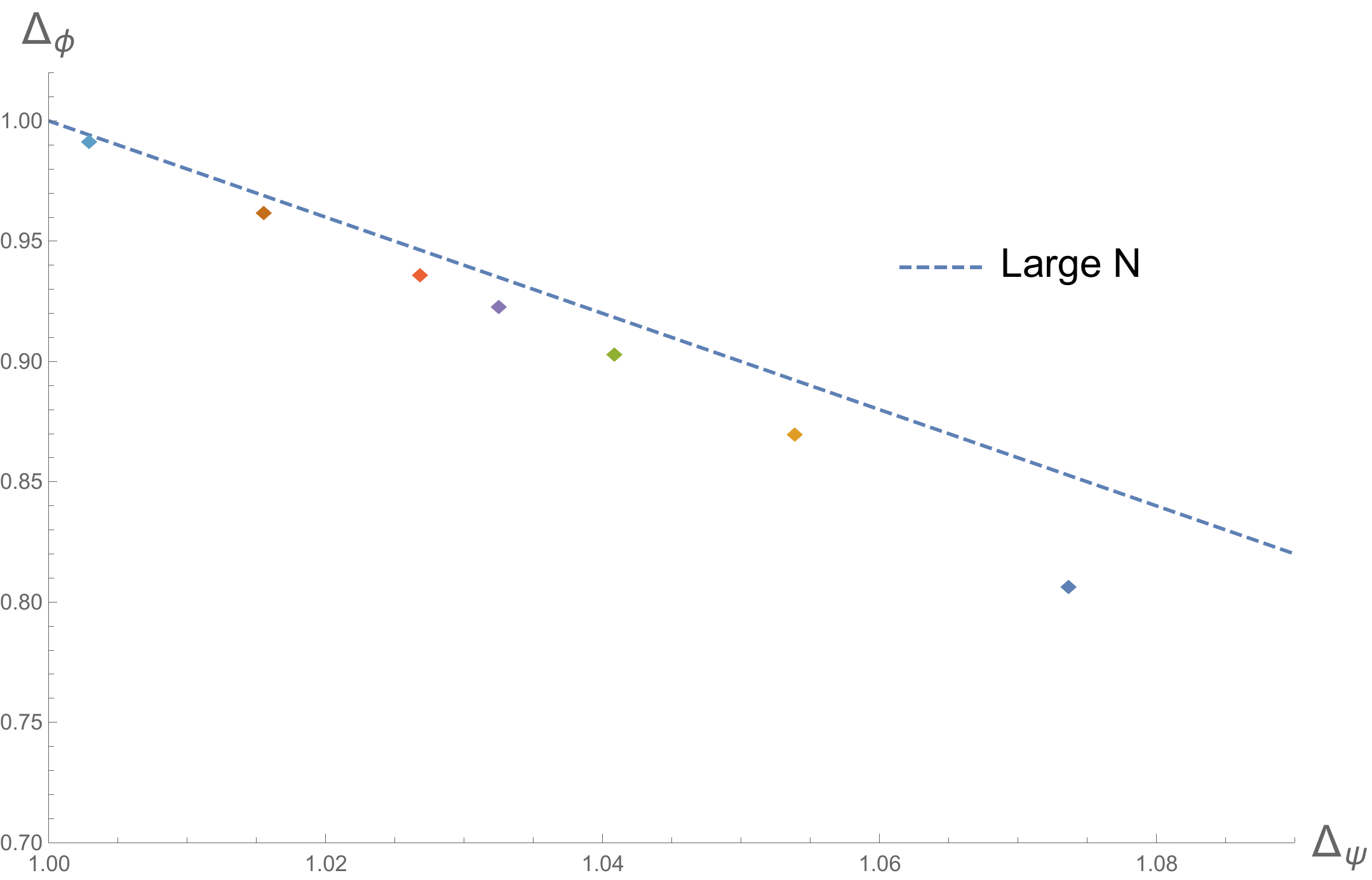}
\caption{Pad\'e estimates in $d=3$ of $\Delta_{\sigma}$ versus $\Delta_{\psi}$ for $N=4,6,8,10,12,20,100$, compared to the large $N$ results 
(\ref{lN-3d-NJL}).}	
\label{PadevsLN-NJL}
\end{figure}

Following the same methods as described in Section \ref{PadeGN}, we now use the $4-\eps$ and $2+\eps$ expansions\footnote{For $\Delta_{\phi}$, we use the 
boundary condition $\Delta_{\phi}=1+{\cal O}(\eps)$ in $d=2+\eps$.} derived in the previous sections to 
obtain rational approximants of scaling dimensions and sphere free energy at the NJL fixed point in $2<d<4$, for $N>2$ (the case $N=2$, 
which displays the emergent supersymmetry, is treated separately in Section \ref{emergentSUSY}). The results in $d=3$ are given in Table \ref{3dNJL-table}, 
indicating which approximants was chosen in each case. In figure \ref{PadevsLN-NJL}, we also plot our 3d estimated 
for $\Delta_{\psi}$ and $\Delta_{\phi}$, compared to the large $N$ results. From the expression for $\Delta_{\phi}$ in (\ref{lN-3d-NJL}), it 
appears that the expansion does not converge well already at ${\cal O}(1/N^2)$, therefore we have just used (\ref{lN-3d-NJL}) to order $1/N$ to produce 
the plot below.  

\begin{table}[h]
\centering
\begin{tabular}{cccccccc}
\hline
\multicolumn{1}{|c|}{$N$}        & \multicolumn{1}{c|}{4} & \multicolumn{1}{c|}{6}& \multicolumn{1}{c|}{8} & \multicolumn{1}{c|}{10} &\multicolumn{1}{c|}{12} &\multicolumn{1}{c|}{20}& \multicolumn{1}{c|}{100} \\ \hline
\multicolumn{1}{|c|}{$\Delta_\psi$ (Pade$_{[3,2]}$)}    & \multicolumn{1}{c|}{1.074}  & \multicolumn{1}{c|}{1.054} & \multicolumn{1}{c|}{1.041}& \multicolumn{1}{c|}{1.033}& \multicolumn{1}{c|}{1.027}& \multicolumn{1}{c|}{1.016}& \multicolumn{1}{c|}{1.0029}  \\ \hline
\multicolumn{1}{|c|}{$\Delta_\phi$ (Pade$_{[2,1]}$)}  & \multicolumn{1}{c|}{0.807} & \multicolumn{1}{c|}{0.870} & \multicolumn{1}{c|}{0.903} &  \multicolumn{1}{c|}{0.923}& \multicolumn{1}{c|}{0.937}&\multicolumn{1}{c|}{0.962}& \multicolumn{1}{c|}{0.992}  \\ \hline
\multicolumn{1}{|c|}{$\Delta_{\phi\bar\phi}$ (Pade$_{[3,1]}$)}      & \multicolumn{1}{c|}{2.018}  & \multicolumn{1}{c|}{2.041} & \multicolumn{1}{c|}{2.055}&  \multicolumn{1}{c|}{2.062}& \multicolumn{1}{c|}{2.064}&\multicolumn{1}{c|}{2.06}  & \multicolumn{1}{c|}{2.022}\\ \hline
\multicolumn{1}{|c|}{$F/(N F_f)$ (Pade$_{[5,2]}$)}  & \multicolumn{1}{c|}{1.109}  & \multicolumn{1}{c|}{1.064} & \multicolumn{1}{c|}{1.045}  &  \multicolumn{1}{c|}{1.034}& \multicolumn{1}{c|}{1.028} & \multicolumn{1}{c|}{1.016} & \multicolumn{1}{c|}{1.0029}  \\ \hline
\multicolumn{1}{l}{}            & \multicolumn{1}{l}{}   & \multicolumn{1}{l}{}   & \multicolumn{1}{l}{}   & \multicolumn{1}{l}{}   & \multicolumn{1}{l}{}  & \multicolumn{1}{l}{}& \multicolumn{1}{l}{}
\end{tabular}
\caption{Estimates of scaling dimensions and sphere free energy at the $d=3$ interacting fixed point of the NJL model.}
\label{3dNJL-table}
\end{table}

For the sphere free energy $\tilde F$, we again  find it convenient to perform the Pad\'e approximation on the quantity 
$f(d) = \tilde F - N \tilde F_{f}$, which corresponds to the interacting part of the NJLY free energy plus the contribution of two free scalars. 
In Figure \ref{F-PadevsLN}, we plot the resulting Pad\'e approximants as a function of $2<d<4$ for a few values of $N$, showing that they approach well the analytical large $N$ formula (\ref{tFLNNJL}).
\begin{figure}[h!]
\centering
\includegraphics[width=8cm]{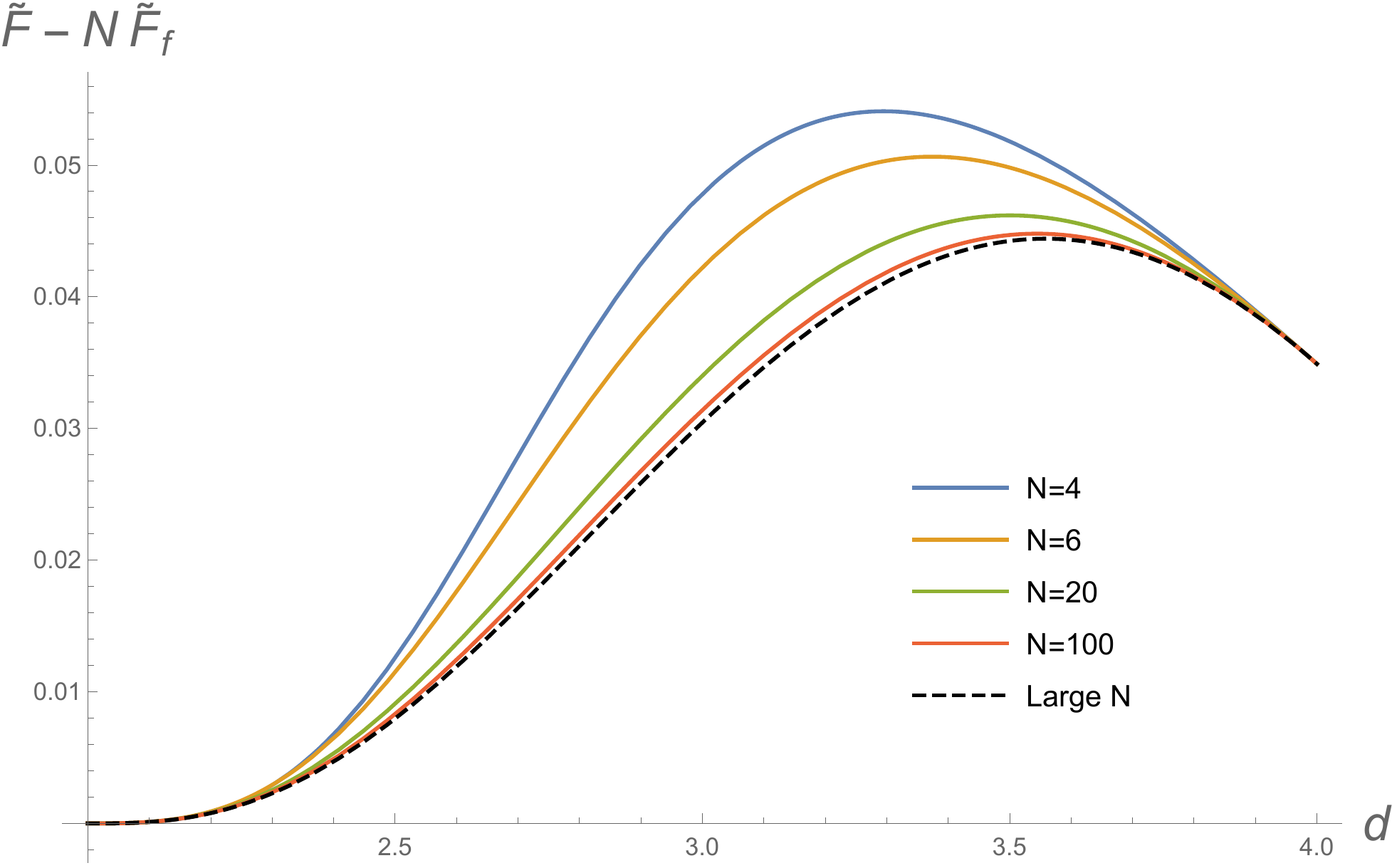}
\caption{Pad\'e estimates of $\tilde{F}-N\tilde F_f$ in $2<d<4$ compared to the large $N$ result (\ref{tFLNNJL}).}
\label{F-PadevsLN-NJL}
\end{figure}
 
We can again use our estimates to test the $d=3$ $F$-theorem and its proposed generalization in $2<d<4$ in terms of $\tilde F$. 
We find that in the whole range of dimensions, the inequalities
\begin{equation}
N \tilde{F}_f + 2\tilde{F}_s > \tilde{F} > N \tilde{F}_f \,
\label{Fth1NJL}
\end{equation}
hold, in accordance with the conjectured generalized $F$-theorem \cite{Giombi:2014xxa,Fei:2015oha}. Using the results in \cite{Fei:2015oha} for the free energy of the $O(2)$ Wilson-Fisher 
model, we have also checked that 
\begin{equation}
N\tilde F_{f}+\tilde F_{O(2)} > \tilde F \ ,
\end{equation}
which is consistent with the fact that the theory can flow from the fixed point consisting of free fermions decoupled from the $O(2)$ model, to the IR stable NJL fixed point.


\section{Models with Emergent Supersymmetry}
\label{emergentSUSY}

\subsection{Theory with 4 supercharges}

A well-known supersymmetric theory with 4 supercharges is the Wess-Zumino model of a single chiral superfield $\Phi$ with superpotential $W\sim \lambda\Phi^3$. 
In $d=4$ the model is classically conformally invariant, but it has a non-vanishing beta function and is expected to be trivial in the IR.  
Continuation of this model to lower dimensions (defined such that the number of supercharges is fixed in $2\le d\le 4$) was discussed in \cite{Giombi:2014xxa,Bobev:2015vsa,Bobev:2015jxa}. In $d=3$, one finds the ${\cal N}=2$ theory of a single chiral superfield 
(a complex scalar and a 2-component Dirac fermion) with cubic superpotential, which flows to a non-trivial CFT in the IR 
\cite{Aharony:1997bx}. In $d=2$, the model matches onto the $(2,2)$ supersymmetric CFT with $c=1$, which is
the $k=1$ member of the superconformal discrete series with $c=\frac {3k}{k+2}$. 

The component Lagrangian of the WZ model in 4d reads
\begin{equation}
\Lagr_{\textrm{WZ}} = \frac{1}{2}(\partial_{\mu}\phi_1)^2+\frac{1}{2}(\partial_{\mu}\phi_2)^2+\frac{1}{2}\bar{\psi}\slashed{\partial}\psi +
\frac{\lambda}{2\sqrt{2}}\bar{\psi}(\phi_1+i\gamma_5 \phi_2)\psi+\frac{\lambda^2}{16}(\bar\phi \phi)^2\,,
\label{WZLag}
\end{equation}
where $\psi$ is a 4-component Majorana fermion and $\phi=\phi_1+i\phi_2$. 
The beta function of the model in $d=4-\eps$ is \cite{Avdeev:1982jx}
\begin{equation}
\beta_{\textrm{WZ}} = -\frac {\eps} {2} \lambda+ \frac{3}{2}\frac{\lambda^3}{(4\pi)^2}-\frac{3}{2}\frac{\lambda^5}{(4\pi)^4}
+\frac{15+ 36 \zeta(3)}{8}\frac{\lambda^7}{ (4\pi)^6}
+{\cal O}(\lambda^9)\,,
\end{equation} 
from which one finds an IR fixed point with $\lambda_*^2 = \frac{16\pi^2\eps}{3}+{\cal O}(\eps^2)$. The dimension of the chiral operator $\phi$ 
at the fixed point is determined by its $R$-charge to be
\begin{equation}
\Delta_{\phi} =\frac{d-1}{2} R_\phi= \frac{d-1}{3}\,.
\label{delphi-WZ}
\end{equation}
One also has the exact result $\Delta_{\phi^2}=\Delta_{\phi}+1$, since the operator $\phi^2$ is obtained from $\phi$ by acting twice 
with the supercharges (this is because, due to cubic superpotential, one has the relation $\Phi^2=0$ in the chiral ring).  
It also follows from supersymmetry \cite{Thomas, Zerf:2016fti} that the dimension of the operator $\phi \bar \phi$ at the fixed point is given by 
\begin{equation}
\Delta_{\phi\bar \phi}= d-2+ \beta'(\lambda_*)= 2-\frac 1 3 \eps^2+ \frac{1+ 12 \zeta(3)}{18} \eps^3 + {\cal O}(\eps^4)\,,
\label{sigsqWZ} 
\end{equation}
which agrees with the explicit three loop calculation of \cite{Zerf:2016fti}. 
In performing Pad\' e extrapolation of this result to $d=3$, we have found that Pad\'e$_{[1,2]}$ and Pad\'e$_{[2,1]}$ give answers close to each other. Their average is
$\approx 1.909$, which is very close to the value $\approx 1.91$ reported using numerical bootstrap studies \cite{Bobev:2015vsa,Bobev:2015jxa}.
We can also take into account the fact that in $d=2$ the dimension of  $\phi \bar \phi$ should approach $2$. Since it also approaches $2$ in $d=4$, it is not a monotonic function of
$d$, which makes Pad\'e extrapolation difficult. If we instead perform a ``two-sided" extrapolation of $\nu^{-1}=d-\Delta_{\phi\bar \phi}$, 
and then return to $\Delta_{\phi\bar \phi}$ in $d=3$,
then we find $\approx 1.94$. This is somewhat further from the numerical bootstrap estimate.

Since in $d=4$ the WZ model includes a complex scalar and a 4-component Majorana fermion (i.e. one half of a Dirac fermion), one would expect it to
correspond to $N_f=1/2$ and $N=2$ in the NJLY model \cite{Thomas,Lee:2006if}. Note that the NJLY Lagrangian (\ref{NJLYlagrange}), specialized to 
the case of a single Majorana fermion, coincides with the WZ Lagrangian (\ref{WZLag}) provided\footnote{One should rescale $\psi \rightarrow \psi/\sqrt{2}$ 
in (\ref{NJLYlagrange}) to get a canonical kinetic term when the fermion is Majorana.}
\begin{equation}
3 g_1^2 =g_2= \frac{3}{2}\lambda^2\,.
\label{SUSY-gs}
\end{equation}
Indeed, setting $N=2$ in the result for the fixed point couplings (\ref{GNYcoupl}), we find:
\begin{equation}
\begin{aligned}
&\frac {(g_{1}^{*})^{2}} {(4\pi)^{2}}=\frac{1}{6}\eps + \frac{1}{18}\eps^2+\mathcal{O}(\eps^3)\,, \\
&\frac{g_{2}^{*}} { (4\pi)^{2}}=\frac{1}{2}\eps + \frac{1}{6}\eps^2+\mathcal{O}(\eps^3)\,.
\end{aligned}
\end{equation}
This is precisely consistent with the relation (\ref{SUSY-gs}), and gives evidence of the emergent supersymmetry in the $N=2$ NJLY model. Note 
that for this value of $N$, the chiral $U(1)$ symmetry of the NJLY model becomes the $U(1)$ $R$-symmetry of the WZ model.\footnote{For a gauge theory in $1+1$ dimensions which exhibits
emergent supersymmetry, see \cite{Gopakumar:2012gd}. In that case a global $U(1)$ symmetry turned into the $U(1)_R$-symmetry of the $(2,2)$ supersymmetric IR CFT.} Further evidence can be found 
by setting $N=2$ in the $4-\epsilon$ expansions of the operator dimensions (\ref{NJLYdim}) and
(\ref{NJLYsigsqcharge}), which give
\begin{align}
\Delta_{\psi} = \frac{3}{2}-\frac{\eps}{3}\ , \qquad
\Delta_{\phi} &= 1-\frac{\eps}{3}\ , \qquad
\Delta_{\phi^2} = 2-\frac{\eps}{3}
\, ,
\end{align}
in agreement with the supersymmetry. In particular, the fact that the $\mathcal{O}(\eps^2)$ terms vanish is consistent with the exact 
result (\ref{delphi-WZ}), and we also see $\Delta_{\phi^2}= \Delta_\phi + 1$ as discussed above. Furthermore, 
setting $N=2$ in (\ref{NJLYsigsq}), we find $\Delta_{\phi\bar \phi}=2-\eps^2/3+ \mathcal{O}(\eps^3)$,
in agreement with the result (\ref{sigsqWZ}) in the WZ model.

It is also interesting to look at the sphere free energy. Setting $N=2$ in the $4-\epsilon$ expansion (\ref{tFNJLY}) of $\tilde F$, we get
\begin{equation}
\tilde{F}_{N=2} =2 \tilde F_s+ 2 \tilde F_f -\frac{\pi \eps^2}{144}  -\frac{\pi \eps^3}{162} +\mathcal{O}(\eps^4)\, .
\label{testofloc}
\end{equation}
This precisely agrees with the expansion of
(5.23) in \cite{Giombi:2014xxa}, which was derived using a proposal for supersymmetric localization in continuous dimension. 
We note that the curvature term (\ref{FbNJLY}) contributes at $\mathcal{O}(\eps^3)$ order to $\tilde F$:  $\delta \tilde{F}_b = - \frac{\pi \eps^3}{1296}$. 
This contribution is crucial for agreement with \cite{Giombi:2014xxa}.
Thus, (\ref{testofloc}) provides a nice perturbative test of the exact formula for $\tilde F$ as a function of $d$ proposed in \cite{Giombi:2014xxa} for the Wess-Zumino model. 

In $d=3$, the result obtained from localization \cite{Jafferis:2010un} yields  $F_{W=\Phi^3} \approx 0.290791$.
In \cite{Fei:2015oha}, the value of $F$ for the $O(2)$ Wilson-Fisher fixed point in $d=3$ was estimated to be $F_{O(2)} \approx 0.124$.
Using also the value $F_f = \frac{1}{8}\log(2)+\frac{3\zeta(3)}{16\pi^2}$ in $d=3$ \cite{Klebanov:2011gs}, 
we see that 
\begin{equation}
2F_f+F_{O(2)} \approx 0.343 > F_{W=\Phi^3}\ ,
\end{equation}
in agreement with the RG flow depicted in Figure \ref{RGs}.

Finally, we discuss $C_T$ of the ${\cal N}=2$ SCFT with superpotential $W\sim \lambda\Phi^3$. Its exact value in $d=3$ has been determined using the supersymmetric localization \cite{Nishioka:2013gza,Witczak-Krempa:2015jca}:
\begin{equation}
\frac {C_T}{C_T^{\rm UV}}= \frac {16(16 \pi - 9 \sqrt 3)}{243 \pi} \approx 0.7268 \ ,
\label{exactCT}
\end{equation}
where $C_T^{\rm UV}= 4 C_{T,s}$ is the value for the free UV theory of two scalars and one two-component Dirac fermion. Let us compare this with a Pad\' e extrapolation of 
the ratio $C_T/C_{T,s}$ using the boundary conditions
\begin{equation}
\frac {C_T}{C_{T,s}} =
\begin{cases} 1  &\mbox{in }\quad d=2\,, \\
5-\frac{11}{6}\eps +\mathcal{O}(\epsilon^{2}) &\mbox{in }\quad d=4-\eps\, ,
\end{cases}
\end{equation}
where the $4-\eps$ expansion was obtained by setting $N=2$ in (\ref{CTNJLY}).
The Pade$_{[1,1]}$ approximant with these boundary conditions is 
\begin{equation}
\frac {C_T}{C_{T,s}} =\frac{49 d-76} {d+20} \ ,
\end{equation}
which in 
$d=3$ gives $C_T/C_{T,s} =71/23\approx 3.087$. Comparing to the free UV CFT, this result implies $C_T^{\rm IR}/C_T^{\rm UV}\approx 0.77$. 
This is not far from the exact result (\ref{exactCT}), demonstrating that the Pad\' e approach works quite well.
It would be 
useful to know the next order in the $4-\eps$ expansion, which may improve the agreement.

\subsection{Theory with 2 supercharges}

It has been suggested that in $d=3$ there exists a minimal ${\cal N}=1$ superconformal theory containing a 2-component Majorana
fermion $\psi$ \cite{Thomas,Grover:2013rc,Bashkirov:2013vya,Iliesiu:2015qra,Shimada:2015gda}. This theory must also contain a pseudoscalar operator $\sigma$, whose scaling dimension 
is related to that of $\psi$ by the supersymmetry, 
\begin{equation}
\Delta_\sigma= \Delta_\psi-\frac{1}{2}
\label{SUSYrel}
\ .
\end{equation}
Some evidence for the existence of this ${\cal N}=1$ supersymmetric CFT was found using the conformal bootstrap \cite{Iliesiu:2015qra}. 

To describe the theory in $d=3$, one can write down the Lagrangian \cite{Thomas,Grover:2013rc,Bashkirov:2013vya, Iliesiu:2015qra}  
\begin{equation}
{\cal L}_{{\cal N}=1} = 
\frac{1}{2}(\partial_{\mu}\sigma)^{2}+\frac{1}{2}\bar{\psi}{\not\,}\partial \psi + \frac{\lambda}{2}\sigma \bar{\psi}\psi
+\frac{\lambda^2}{8}\sigma^4\,.
\label{L-cN1}
\end{equation}
This model has ${\cal N}=1$ supersymmetry in $d=3$; the field content can be packaged in the real superfield 
$\Sigma = \sigma+\bar\theta \psi+\frac{1}{2}\bar\theta\theta f$, and the interactions follow from the cubic superpotential 
$W\sim \lambda \Sigma^3$.\footnote{To obtain the component Lagrangian (\ref{L-cN1}), one should eliminate the auxiliary field $f$ using 
its equation of motion $f \sim \lambda \sigma^2$.}
It is natural to expect that this model flows to a non-trivial ${\cal N}=1$ SCFT in the IR. Note that the theory cannot be described 
as the UV fixed point of a lagrangian for a Majorana fermion with quartic interaction, because the term $(\bar \psi \psi)^2$ vanishes for a 2-component spinor. 

The theory (\ref{L-cN1}) is super-renormalizable in $d=3$, and one may attempt its $4-\epsilon$ expansion \cite{Thomas}. To formulate a Yukawa theory in $d=4$, 
one strictly speaking needs a 4-component Majorana fermion, which corresponds to the GNY model with $N=2$. However, the GNY description may be formally continued
to $N=1$. A sign of the simplification that occurs for this value is that $\sqrt{N^2+ 132 N + 36}$, which appears in the $4-\epsilon$ expansions (\ref{GNYdims}), (\ref{GNYsigmasq}), equals 13 for $N=1$. For this value of $N$, we find that the fixed point couplings in (\ref{GNYcoupl}) become
\begin{equation}
\begin{aligned}
&\frac {(g_{1}^{*})^{2}} {(4\pi)^{2}}=\frac{1}{7}\eps + \frac{3}{49}\eps^2+\mathcal{O}(\eps^3)\,,\\
& \frac{g_{2}^{*}} { (4\pi)^{2}}=\frac{3}{7}\eps + \frac{9}{49}\eps^2+\mathcal{O}(\eps^3)\,.
\label{N1-GNY-gs}
\end{aligned}
\end{equation}
This is consistent with the exact relation $3g_1^2=g_2$ in the SUSY model. Indeed the GNY Lagrangian (\ref{GNY-Lag}), formally applied to the 
case of a single 2-component Majorana fermion, coincides with (\ref{L-cN1}) when $3g_1^2=g_2 = 3\lambda^2$. The result (\ref{N1-GNY-gs}) 
gives a two-loop evidence that the non-supersymmetric GNY model with $N=1$ flows at low energies to a supersymmetric fixed point.

In \cite{Thomas} it was found using one-loop calculations that $\Delta_\sigma = 1- \frac {3\epsilon}{7}=  \Delta_\psi-\frac{1}{2}$.
Let us check that the supersymmetry relation (\ref{SUSYrel}) continues to hold at order $\epsilon^2$. Using (\ref{GNYdims}) with $N=1$, we indeed find
\begin{align}
\Delta_{\sigma} &= 1-\frac{3}{7}\eps +\frac{1}{49} \eps^{2}+ \mathcal{O}(\eps^3)\,,\\
\Delta_{\psi} &= \frac{3}{2}-\frac{3}{7}\eps +\frac{1}{49} \eps^{2} +\mathcal{O}(\eps^3)\,.
\end{align}
The dimensions of operators $\sigma^2$ and $\sigma \psi$ should also be related by the supersymmetry, $\Delta_{\sigma^2}= \Delta_{\sigma \psi}-\frac{1}{2}$.
Since $\sigma \psi$ is a descendant, we also have $\Delta_{\sigma \psi}= \Delta_\psi+1$. Thus, the supersymmetry relation assumes the form \cite{Bashkirov:2013vya}
\begin{equation}
\Delta_{\sigma^2}= \Delta_{\psi}+\frac{1}{2}= \Delta_\sigma+1\ .
\label{anotherSUSYrel}
\end{equation}
Substituting $N=1$ into (\ref{GNYsigmasq}) we find
\begin{align}
\Delta_{\sigma^2} &= 2-\frac{3}{7} \eps +\frac{1}{49} \eps^{2}+ \mathcal{O}(\eps^3)\, ,
\end{align}
so that the supersymmetry relation (\ref{anotherSUSYrel}) holds to order $\eps^2$.
These non-trivial checks provide strong evidence that the continuation of the GNY model to $N=1$ flows 
to a superconformal theory to all orders in the $4-\epsilon$ expansion.

If we apply the standard  Pad\' e$_{[1,1]}$ extrapolation, we find
\begin{equation}
\Delta_\sigma= \frac{ 8d-11}{25-d}\ ,
\label{oneoneP}
\end{equation}
which in $d=3$ gives $\Delta_\sigma= \frac{13}{22}\approx 0.59$. \footnote{We note that our estimate $\Delta_{\sigma^2}\approx 1.59$ in the $N=1$ theory is below 2, just like the estimate (\ref{ntwoest}) in the $N=2$ theory. This is in contrast with the large $N$
behavior (\ref{lN-3d}) where $\Delta_{\sigma^2}>2$. Thus, perhaps not surprisingly, the theories with $N=1, 2$ are, in some respects, rather far from the large $N$ limit.}
This is close to the estimate of $\Delta_{\sigma}$ obtained using the numerical bootstrap \cite{Iliesiu:2015qra}. It is the value $\Delta_\sigma\approx 0.582$ 
where the boundary of the excluded region touches the SUSY line $\Delta_{\sigma}=\Delta_{\psi}-1/2$.

\begin{figure}[h!]
\centering
\includegraphics[width=10cm]{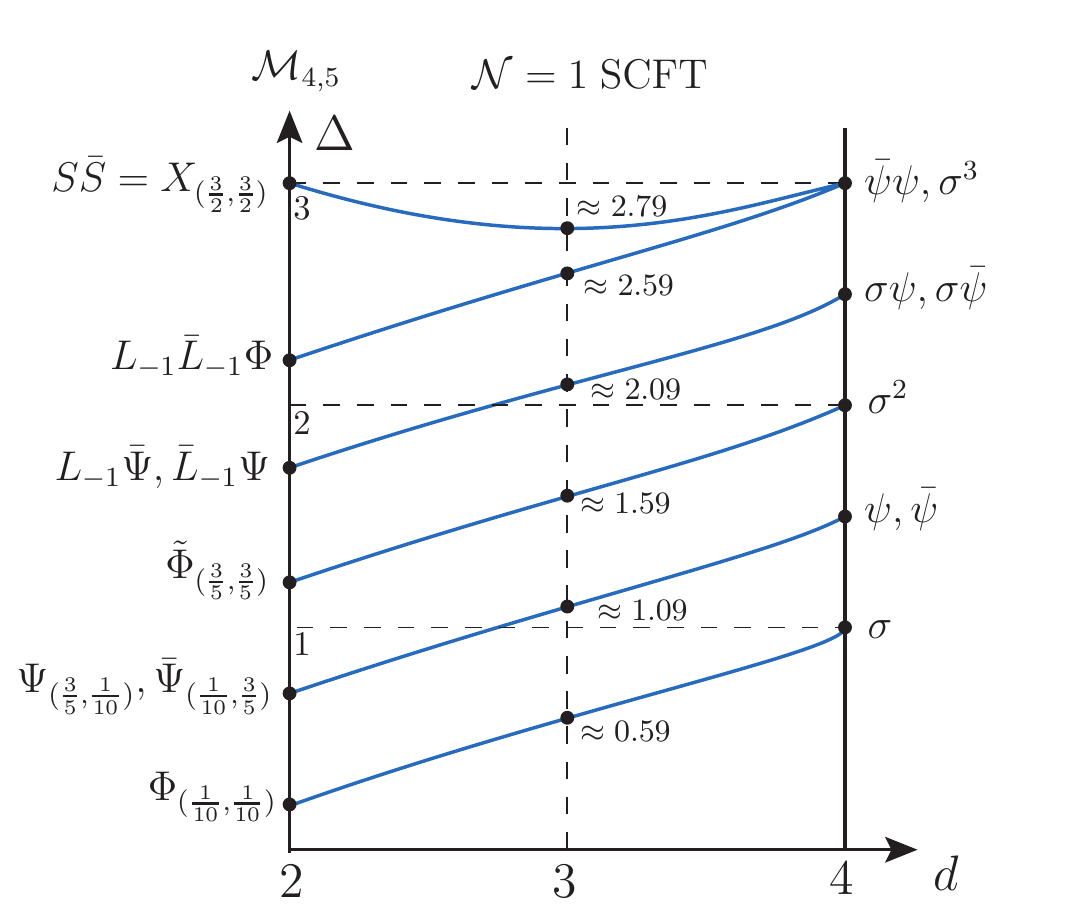}
\caption{Qualitative picture of the interpolation of operator dimensions from the $d=4$ free theory to the tri-critical Ising model ${\cal M}(4,5)$ in $d=2$, indicating our estimated values for the ${\cal N}=1$ SCFT in $d=3$. In $2\le d <4$, the operators $\sigma\psi$, and a linear combination of $\sigma^3$ and $\bar\psi \psi$, are expected to be conformal descendants of $\psi$ and $\sigma$ respectively.}
\label{tricritmatch}
\end{figure}

It is also important to know how the theory behaves when continued to $d=2$. It is plausible that the $d=2$ theory has ${\cal N}=1$ superconformal symmetry, and
the obvious candidate is the tri-critical Ising model \cite{Grover:2013rc,Shimada:2015gda}, which is the 
simplest supersymmetric minimal model \cite{Belavin:1984vu,Friedan:1984rv}. The Pad\' e extrapolation (\ref{oneoneP}) gives $\Delta_\sigma\approx 0.217$,
which is quite close to the dimension $1/5$ of the energy operator in the tri-critical Ising model. This provides new evidence that the GNY model with $N=1$ extrapolates to the tri-critical Ising model in $d=2$; in figure \ref{tricritmatch} we show how the operator spectrum matches with the exact results in $d=2$.
Imposing the boundary condition that $\Delta_\sigma=1/5$ in $d=2$ enables us to perform a ``two-sided" Pad\' e estimate. The resulting value in $d=3$ is
$\approx 0.588$, which is very close to that following from (\ref{oneoneP}). The agreement with the bootstrap result $\Delta_\sigma\approx 0.582$ is excellent.   

From the $\beta$-functions (\ref{betasGNY}) for the GNY model we may also deduce the dimensions of two primary operators that are mixtures of $\sigma^4$ and $\sigma \bar\psi\psi$.
They are determined by the eigenvalues $\lambda_1, \lambda_2$ of the matrix $\partial \beta_i/\partial g_j$. At the fixed point for $N=1$ we find
\begin{align}
\Delta_{1} &= d+\lambda_1= 4 - \frac{3}{7} \eps^2+ \mathcal{O}(\eps^3)\,, \notag \\
\Delta_{2} &= d+\lambda_2= 4 +  \frac{6}{7}  \eps - \frac{95}{49} \eps^2+ \mathcal{O}(\eps^3)\,, 
\label{GNYmarg}
\end{align}
Since $\Delta_1$ corresponds to the $\theta\bar\theta$ component of the superfield $\Sigma^3$, we know that 
\begin{equation}
\Delta_{\sigma^3}= \Delta_1-1= 3 - \frac{3}{7} \eps^2+ \mathcal{O}(\eps^3)\ .
\label{cubicest}
\end{equation}
As a check, we may set $N=1$ in (\ref{GNYpsipsisigma3ep}) and see that the order $\eps$ term in $\Delta_{\sigma^3}$ indeed vanishes. 
It is further possible to argue that this primary operator, when continued to $d=2$, matches onto the dimension 3 operator in the tri-critical Ising model, i.e. the product of left and right supercurrents.
This implies that $ \Delta_{\sigma^3}$ is not monotonic as a function of $d$ and is likely to be somewhat smaller than 3 in $d=3$.\footnote{This is in line with a statement
in \cite{Iliesiu:2015qra} that a kink lies on the SUSY line when $ \Delta_{\sigma^3}$ is slightly smaller than 3.}
Simply setting $\eps=1$ in (\ref{cubicest}) gives $\Delta_{\sigma^3}\approx 2.57$, but this probably underestimates it. One way to implement the exact boundary condition $\Delta_{\sigma^3}=3$ in $d=2$ is to extrapolate the quantity $\tilde{\Delta} = (\Delta_{\sigma^3}-3)/(d-2)$, instead of $\Delta_{\sigma^3}$ itself \cite{Guida:1998bx}. 
This approach would yield 
the estimate $\Delta_{\sigma^3}\approx 2.79$ in $d=3$. 
This slightly relevant operator is parity odd
in $d=3$, while the dimension $\Delta_1$ corresponds to an irrelevant parity even operator.

\begin{figure}[h!]
\centering
\includegraphics[width=10cm]{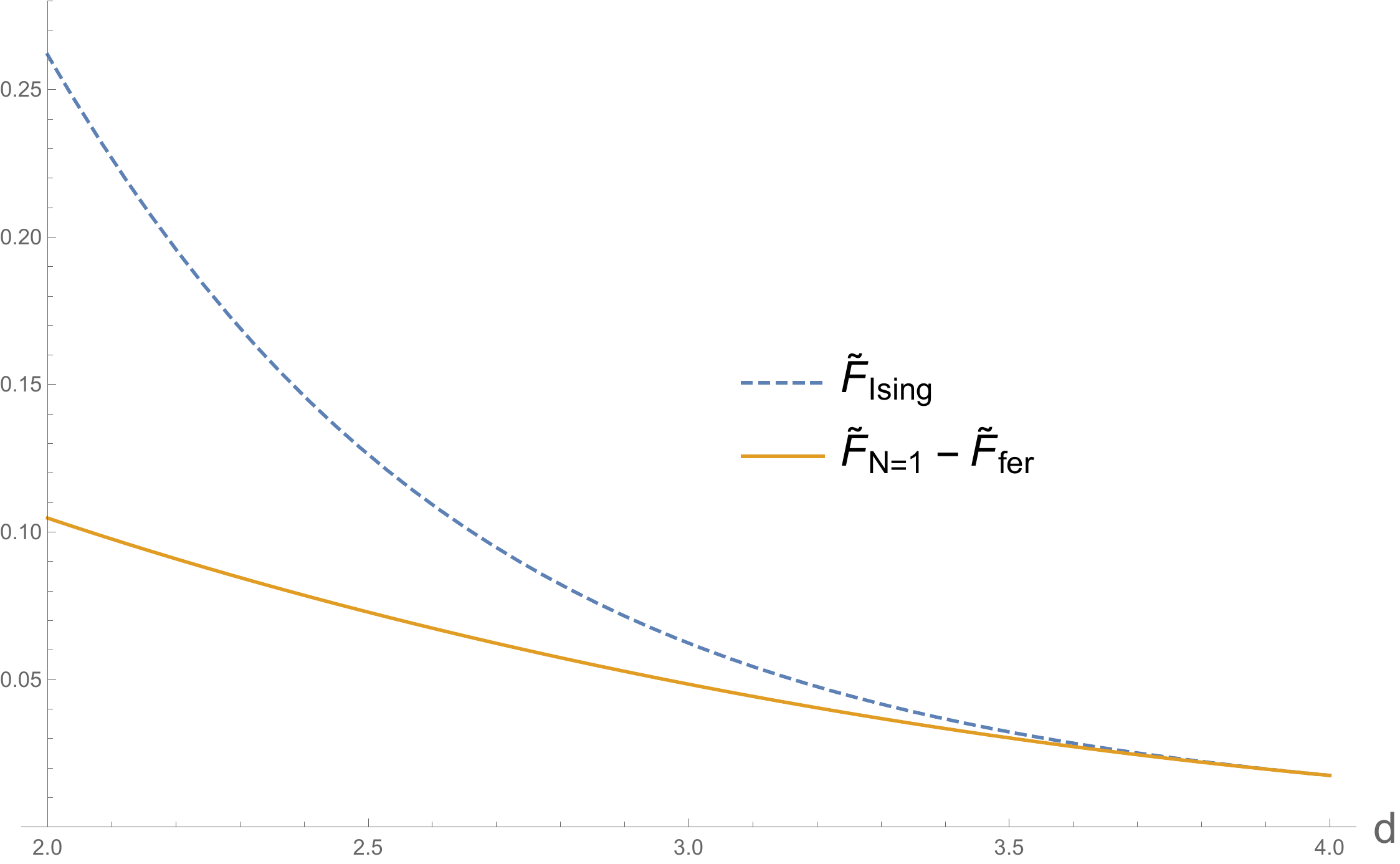}
\caption{Comparison of $\tilde F-\tilde F_f$ at the SUSY fixed point and $\tilde F_{\rm Ising}$ (from \cite{Fei:2015oha}) in $2 \le d \le 4$.}
\label{FGNYvIsing}
\end{figure}

Similarly, we can perform extrapolation of $\tilde F$ and check if the $d=2$ value is close to ${\pi c/6}$ where $c=7/10$ is the central charge of the tri-critical Ising model.
Setting $N=1$ in  (\ref{tFGNYall}), we have
\begin{align}
&\tilde{F}=\tilde F_s+ \tilde F_f -\frac {\pi}{6} \left ( \frac{\epsilon^2}{112} + \frac{\epsilon^3}{98} +\mathcal{O} (\epsilon^{4}) \right).
\end{align}
Performing a Pad\' e approximation of the quantity $f(d)=\tilde{F} - \tilde{F}_{f}$, we find that the average of the standard Pad\'e$_{[2,1]}$ and Pad\'e$_{[1,2]}$ approximants 
yields  $\tilde F/\tilde F_s \approx 0.68$, which is quite close to $c=7/10$. Therefore, in order to get a better 
estimate in $d=3$, it makes sense to impose it as an exact boundary condition in $d=2$. Following this procedure, and taking an average of the Pad\'e approximants 
with $n+m=4$, we find the $d=3$ estimate
\begin{equation}
F \approx 0.158\,.
\label{FSUSY}
\end{equation}
In the UV, we have the free CFT of one scalar and one Majorana fermion, which has $F^{\rm UV} =\frac{1}{4}\log 2 \approx  0.173$, and therefore we find $F^{\rm IR}/F^{\rm UV}\approx 0.91$. This is a check of the $F$-theorem for the flow from the free to the interacting ${\cal N}=1$ SCFT. It is also interesting to compare the value of $F$ at the SUSY fixed point to the decoupled Ising fixed point in Figure \ref{RGs}. A plot comparing $\tilde F-\tilde F_f$ with $\tilde F_{\rm Ising}$, which was obtained in \cite{Fei:2015oha}, is given in Figure \ref{FGNYvIsing}. It shows that $\tilde F <\tilde F_f+\tilde F_{\rm Ising}$ in the whole range $2<d<4$, in agreement with the generalized $F$-theorem 
\cite{Giombi:2014xxa,Fei:2015oha} and the expectation that the SUSY fixed point is IR stable. 

Finally we consider $C_T$. Its $4-\eps$-expansion for general $N$ is given in eq.~(\ref{CT-GNY}), and we can use it to
estimate the $C_T$ value at the ${\cal N}=1$ SUSY fixed point in $d=3$. We can perform a Pad\' e approximation on the ratio $C_T/C_{T,s}$, where $C_{T,s}$ is the free scalar value, using the boundary conditions
\begin{equation}
\frac {C_T}{C_{T,s}} =
\begin{cases} \frac{7}{10}  &\mbox{in }\quad d=2\,, \\
\frac{5}{2}-\frac{19}{28}\eps +\mathcal{O}(\epsilon^{2}) &\mbox{in }\quad d=4-\eps\,,
\end{cases}
\end{equation}
where we have imposed the $d=2$ value corresponding to the tri-critical Ising model. A Pade$_{[1,1]}$ approximant with these boundary conditions is 
\begin{equation}
\frac {C_T}{C_{T,s}} =\frac{497 d-728} {62 d+256} \ ,
\end{equation}
which in 
$d=3$ gives $C_T/C_{T,s} \approx 1.73$. This value lies well within the region allowed by the bootstrap in fig. 8 of \cite{Iliesiu:2015qra} (note that $\Delta_\psi\approx 1.09$). Comparing to the free UV CFT of one scalar and one Majorana fermion, we have found $C_T^{\rm IR}/C_T^{\rm UV}\approx 0.86$. It would be 
useful to know the next order in the $4-\eps$ expansion in order to obtain a more precise estimate. Based on the comparison of the Pad\' e for the ${\cal N}=2$ model
with the exact result, we may expect the $\eps^2$ correction to reduce $C_T$ somewhat in $d=3$.

Let us also include a brief discussion of the non-supersymmetric fixed point of the $N=1$ GNY model, marked by the red triangle in figure \ref{RGs}.
While it has $g_2^* < 0$ in $d=4-\eps$, it may become stable for sufficiently small $d$. Changing the sign of the square root in
(\ref{tFGNYall}), we find the $4-\eps$ expansion
of the sphere free energy at this fixed point: 
\begin{align}
&{\tilde F}_{\rm non-SUSY}=\tilde F_s+ \tilde F_f -\frac {\pi}{6} \left ( \frac{\epsilon^2}{112} + \frac{ 6875 \epsilon^3}{889056} +\mathcal{O} (\epsilon^{4}) \right)\ .
\label{nonSUSYF}
\end{align}
Extrapolating this to $d=2$, we estimate $c_{\rm non-SUSY}\approx 0.78$. This is close to central charge $4/5$ of the $(5,6)$ minimal model.
This model includes primary fields of conformal weight $1/40$ and $21/40$, so that one may form a spin $1/2$ field with weights $(h,\tilde h)=(1/40,21/40)$ that could correspond
to $\psi$ in the Yukawa theory (such a field is not present in the standard modular invariants that retain fields of integer spin only).\footnote{
Another possibility is that the continuation of the non-supersymmetric fixed point to $d=2$ gives the $(6,7)$ minimal model
(we are grateful to the referee for suggesting this). Its central charge $6/7\approx 0.857$ is also not far from the extrapolation of (\ref{nonSUSYF}) to $d=2$.
The spectrum of the $(6,7)$ model contains a spin $1/2$ operator with weights  $(h,\tilde h)=(5/56, 33/56)$.}
In $d=3$ 
our extrapolation gives $F_{\rm non-SUSY}\approx 0.16$. The latter quantity is bigger than (\ref{FSUSY}); therefore, 
the non-SUSY fixed point, if it is stable, can flow to the SUSY one in $d=3$. 

It is also interesting to look at the scaling dimensions of $\sigma$, $\psi$ and $\sigma^2$ at the non-supersymmetric fixed point. 
Changing the sign of the square roots in (\ref{GNYdims}) and (\ref{GNYsigma2}), we find for $N=1$:
\begin{equation}
\begin{aligned}
& \Delta_{\sigma}^{\rm non-SUSY} = 1 - \frac{3}{7}\eps - \frac{95}{6174}\eps^2 + \mathcal{O}(\eps^3)\, ,\\
& \Delta_{\psi}^{\rm non-SUSY} = \frac{3}{2} - \frac{3}{7}\eps - \frac{115}{37044}\eps^2  + \mathcal{O}(\eps^3)\, , \\
& \Delta_{\sigma^2}^{\rm non-SUSY} = 2 - \frac{22}{21}\eps + \frac{1117}{9261}\eps^2 + \mathcal{O}(\eps^3)\,
\label{nonSUSYdims}
\end{aligned}
\end{equation}
Using Pade$_{[1,1]}$ extrapolations to $d=2$ we get 
\begin{equation}
\Delta_{\sigma}^{\rm non-SUSY} \approx 0.077\ , \qquad
\Delta_{\psi}^{\rm non-SUSY} \approx 0.63\ , \qquad  
\Delta_{\sigma^2}^{\rm non-SUSY} \approx 0.297\ .
\end{equation}
These numbers are not far from the corresponding operator dimensions in either the $(5,6)$ or the $(6,7)$ minimal models.
For example, in the $(5,6)$ interpretation the exact scaling dimension of $\psi$ in $d=2$ should be $11/20$, while in the $(6,7)$ it should be $19/28$. 
The Pad\' e value we find lies between the two,
but the accuracy of the extrapolation all the way to $d=2$ is hard to assess.

The Pade$_{[1,1]}$ extrapolations of (\ref{nonSUSYdims}) to $d=3$ yield $\Delta_{\sigma}^{\rm non-SUSY} \approx 0.555$ and $\Delta_{\psi}^{\rm non-SUSY} \approx 1.068$.  Interestingly, 
these values are not far from the feature at $(\Delta_{\sigma},\Delta_{\psi}) \approx  
(0.565, 1.078)$, which lies below the supersymmetry line in figure 7 of \cite{Iliesiu:2015qra}.  Changing the sign of the square root in (\ref{GNYpsipsisigma3ep}), we also find
\begin{equation}
\Delta_{\sigma^3}^{\rm non-SUSY}= 3 - \frac{13}{7} \eps + \mathcal{O}(\eps^2)\, ,
\end{equation}
which suggests that in $d=3$ the dimension of this parity-odd operator is less than 3. A higher loop analysis of the operator dimensions is, of course, desirable. 

\section*{Acknowledgments}

We thank David Gross, David Poland, Silviu Pufu and David Simmons-Duffin for useful discussions. 
IRK and GT gratefully acknowledge support from the Simons Center for Geometry and Physics, Stony Brook University 
at which some of this research was performed.
This work was supported in part by
the US NSF under Grants No. PHY-1314198, PHY-1318681, PHY-1620059 and PHY-1620542.


\appendix
\section{Beta functions  and anomalous dimensions  for general Yukawa theories}
In this appendix  we list known general results for  $\beta$- and $\gamma$-functions for general Yukawa theories in $d=4$ with the Lagrangian 
\begin{align}
\mathcal{L} = \frac{1}{2}(\partial_{\mu}\phi_{i})^{2} + \bar{\psi} \slashed{\partial}\psi +\bar{\psi}\Gamma_{i}\psi \phi_{i} +\frac{1}{4!}g_{ijkl}\phi_{i}\phi_{j}\phi_{k}\phi_{l}\,,
\end{align}
where $\phi_{i}$, with $i=1,\dots, N_{b}$ are real scalar fields,  $\psi_{\alpha}$, with $\alpha=1,\dots, N_{f}$ are  four-component Dirac spinors, and the matrices $\Gamma_{i}$ have the following general form 
\begin{align}
\Gamma_{i} = S_{i}\otimes \bold{1} + iP_{i} \otimes \gamma_{5}, \quad \Gamma_{i}^{\dag} = S_{i}^{\dag}\otimes \bold{1} - iP^{\dag}_{i} \otimes \gamma_{5}
\end{align}
and act in the flavor and spinor spaces and are not necessarily Hermitian.  We also assume that $\gamma_{5}^{2}=\bold{1}$.

Using the papers \cite{Machacek:1983tz, Machacek:1983fi, Machacek:1984zw} and   \cite{Jack:1990eb} one can find  the $\beta$- and $\gamma$-functions 
of the general Yukawa theory\. footnote{We note that one can use results of  \cite{Machacek:1983tz, Machacek:1983fi, Machacek:1984zw} for the four-component spinors, see sec. 4 in \cite{Machacek:1983fi}. } For the $\gamma$-functions the result reads  (see formulas (3.6), (4.4) in \cite{Machacek:1983tz} and (7.2) in  \cite{Jack:1990eb}) 
\begin{align}
&\gamma_{\psi} = \frac{1}{(4\pi)^{2}}\frac{1}{2}\Gamma_{i}\Gamma_{i}^{\dag}-\frac{1}{(4\pi)^{4}}\Big(\frac{1}{8}\Gamma_{i}\Gamma_{j}^{\dag}\Gamma_{j}\Gamma_{i}^{\dag}+\frac{3}{8}\tr(\Gamma_{i}\Gamma_{j}^{\dag})\Gamma_{i}^{\dag}\Gamma_{j}\Big)\,,\notag\\
&\gamma_{\phi, ij}= \frac{1}{2(4\pi)^{2}}\tr(\Gamma_{i}\Gamma^{\dag}_{j})+ \frac{1}{(4\pi)^{4}}\Big(\frac{1}{12}g_{iklm}g_{jklm}-\frac{3}{4}\tr(\Gamma_{j}\Gamma_{i}^{\dag}\Gamma_{k}\Gamma_{k}^{\dag}) - \frac{1}{2}\tr(\Gamma_{j} \Gamma_{k}^{\dag}\Gamma_{i}\Gamma_{k}^{\dag})\Big)\,.
\label{Gamma_generalYukawa}
\end{align}
For the  anomalous  mixing matrix  of the $N_{b}(N_{b}+1)/2$ quadratic operators $O_{ij}=\phi_{i}\phi_{j}$ with $i\leqslant j$ we have \cite{Pernici:1999nw}
\begin{align}
\gamma_{ij,kl} = \gamma_{\phi, mk }\delta_{ij,ml}+\gamma_{\phi, ml}\delta_{ij,km}+\begin{cases}
M_{ij,kl}+M_{ij,lk}\,,\quad k\neq l\,, \\
M_{ij,kk}\,, ~~~~~~~~~~~~~ k=l\,,
\end{cases}\quad (i\leqslant j, k\leqslant l)\,,
\end{align}
where $\delta_{ij,kl}$ is the Kronecker delta  (e.g. $\delta_{11,11}=1, \delta_{11,12}=0$, $\delta_{12,12}=1$, $\delta_{22,22}=1$,\dots )  and 
\begin{align}
M_{ij,kl}=\frac{1}{(4\pi)^{2}} g_{ijkl} - \frac{1}{(4\pi)^{4}}\Big(g_{ikmn}g_{jlmn}+\tr(\Gamma_{l}\Gamma_{m}^{\dag})g_{ijkm}-2\tr(\Gamma_{i}\Gamma_{k}^{\dag}\Gamma_{j}\Gamma_{l}^{\dag})\Big)\,.
\end{align}
For the $\beta$-functions we have (see  (3.3)  in \cite{Machacek:1983fi} and (7.2) in \cite{Jack:1990eb})
\begin{align}
\beta_{i} =& \frac{1}{(4\pi)^{2}}\Big(\frac{1}{2}(\Gamma^{2\dag}\Gamma_{i}+\Gamma_{i}\Gamma^{2})+2\Gamma_{j}\Gamma_{i}^{\dag}\Gamma_{j}+\frac{1}{2}\Gamma_{j}\tr(\Gamma_{j}^{\dag}\Gamma_{i})\Big)+\frac{1}{(4\pi)^{4}}\Big(2\Gamma_{k}\Gamma_{j}^{\dag}\Gamma_{i}(\Gamma_{k}^{\dag}\Gamma_{j}-\Gamma_{j}^{\dag}\Gamma_{k}) \notag\\
&-\Gamma_{j}(\Gamma^{2}\Gamma_{i}^{\dag}+\Gamma_{i}^{\dag}\Gamma^{2\dag})\Gamma_{j}-\frac{1}{8}(\Gamma_{j}\Gamma^{2}\Gamma_{j}^{\dag}\Gamma_{i}+\Gamma_{i}\Gamma_{j}^{\dag}\Gamma^{2\dag}\Gamma_{j})-\tr(\Gamma_{i}^{\dag}\Gamma_{k})\Gamma_{j}\Gamma_{k}^{\dag}\Gamma_{j}\notag\\
&-\frac{3}{8}\tr(\Gamma_{j}^{\dag}\Gamma_{k})(\Gamma_{j}\Gamma_{k}^{\dag}\Gamma_{i}+\Gamma_{i}\Gamma_{k}^{\dag}\Gamma_{j})-\Gamma_{j}\tr\big(\frac{3}{8}(\Gamma^{2}\Gamma_{j}^{\dag}+\Gamma_{j}^{\dag}\Gamma^{2\dag})\Gamma_{i}+\frac{1}{2}\Gamma_{j}^{\dag}\Gamma_{k}\Gamma_{i}^{\dag}\Gamma_{k}\big) \notag\\
&-2g_{ijkl}\Gamma_{j}\Gamma_{k}^{\dag}\Gamma_{l}+\frac{1}{12}g_{iklm}g_{jklm}\Gamma_{j}\Big)
\end{align}
and (see  (4.3)  in \cite{Machacek:1984zw})
\begin{align}
\beta_{ijkl}  =& \frac{1}{(4\pi)^{2}}\Big(\frac{1}{8}\sum_{\textrm{perm}}g_{ijmn}g_{mnkl}-\frac{1}{2}\sum_{\textrm{perm}}\tr(\Gamma_{i}\Gamma^{\dag}_{j}\Gamma_{k}\Gamma_{l}^{\dag})+\frac{1}{2}\sum_{a=i,j,k,l}\tr(\Gamma^{\dag}_{a}\Gamma_{a})g_{ijkl}\Big)\notag\\
&+\frac{1}{(4\pi)^{4}}\Big( \frac{1}{12}g_{ijkl}\sum_{a=ijkl}g_{amnp}g_{amnp}- \frac{1}{4}\sum_{\textrm{perm}}g_{ijmn}g_{kmpr}g_{lnpr}-\frac{1}{8}\sum_{\textrm{perm}}\Tr(\Gamma_{n}^{\dag}\Gamma_{p})g_{ijmn}g_{klmp}\notag\\
&+\frac{1}{2}\sum_{\textrm{perm}}g_{ijmn}\tr(\Gamma_{k}\Gamma_{m}^{\dag}\Gamma_{l}\Gamma^{\dag}_{n})- g_{ijkl}\sum_{a=ijkl}\big(\frac{3}{4}\tr(\Gamma_{a}\Gamma^{\dag}_{a}\Gamma_{m}\Gamma_{m}^{\dag})+\frac{1}{2}\tr(\Gamma_{a}\Gamma_{m}^{\dag}\Gamma_{a}\Gamma_{m}^{\dag})\big) \notag\\
&+\sum_{\textrm{perm}}\big(\tr(\Gamma_{m}^{\dag}\Gamma_{m}\Gamma^{\dag}_{i}\Gamma_{j}\Gamma^{\dag}_{k}\Gamma_{l})+2 \tr(\Gamma_{m}\Gamma^{\dag}_{i}\Gamma_{m}\Gamma^{\dag}_{j}\Gamma_{k}\Gamma^{\dag}_{l})+
\tr(\Gamma_{i}\Gamma^{\dag}_{j}\Gamma_{m}\Gamma^{\dag}_{k}\Gamma_{l}\Gamma^{\dag}_{m})\big) \Big)\,,
\end{align}
where $\sum_{\textrm{perm}}$ denotes the sum over all permutation of the indices $i,j,k,l$ and $\Gamma^{2}=\Gamma_{i}^{\dag}\Gamma_{i}$ and $\Gamma^{2\dag}=\Gamma_{i}\Gamma_{i}^{\dag}$, also $\sum_{a=ijkl} f(a) \equiv f(i)+f(j)+f(k)+f(l) $. Traces $\tr (\Gamma_{i}\dots)$ are  over the flavor and spinor indices and $\tr \bold{1}=4$, $\tr \gamma_{5}=0$.

Looking at the previous expressions for the $\beta$- and $\gamma$-functions, one can easily generalize the result of \cite{Jack:1990eb} for $\beta_{b}$, when the matrices $\Gamma_{i}$ are not 
Hermitian:
\begin{align}
&\beta_{b} = -\frac{1}{(4\pi)^{8}} \frac{1}{144}\Big(\frac{1}{8}\tr(\Gamma_{i}^{\dagger}\Gamma^{2}\Gamma_{i}\Gamma^{2})+\tr(\Gamma_{i}^{\dag}\Gamma_{j}\Gamma_{i}^{\dag}\Gamma_{j}\Gamma^{2})+\tr(\Gamma_{i}\Gamma_{j}^{\dag}\Gamma_{i}\Gamma_{k}^{\dag}\Gamma_{j}\Gamma^{\dag}_{k})-\tr(\Gamma_{i}\Gamma_{j}^{\dag}\Gamma_{k}\Gamma_{i}^{\dag}\Gamma_{j}\Gamma_{k}^{\dag})\notag\\
&~+\frac{3}{4} \tr(\Gamma_{i}^{\dag}\Gamma_{j})\tr(\Gamma_{i}^{\dag}\Gamma_{j}\Gamma^{2}+\Gamma_{i}^{\dag}\Gamma_{k}\Gamma_{j}^{\dag}\Gamma_{k})+g_{ijkl}\tr(\Gamma_{i}\Gamma_{j}^{\dag}\Gamma_{k}\Gamma_{l}^{\dag})-\frac{1}{24}g_{iklm}g_{jklm} \tr(\Gamma_{i}^{\dag}\Gamma_{j})\Big)\,.
\end{align}
Now to use these formulas for the GNY model  one simply takes
\begin{align}
\Gamma_{1}= g_{1} \mathbbm{1}_{N_{f}\times N_{f}} \otimes \bold{1}, \quad g_{1111}=g_{2}\,.
\end{align}
and  finds the results  (\ref{betasGNY}), (\ref{GNYdim}),  (\ref{GNYsigma2}) and (\ref{betabGNY}). In the case of the NJL model, we have 
\begin{align}
&\Gamma_{1}= g_{1} \mathbbm{1}_{N_{f}\times N_{f}} \otimes \bold{1}, \quad \Gamma_{2}=ig_{1}\mathbbm{1}_{N_{f}\times N_{f}} \otimes \gamma_{5}\,, \notag\\
&g_{1111}=g_{2222}=g_{2}, \quad g_{1122}=g_{1221}=\dots =\frac{1}{3}g_{2}\,,
\end{align}
and we obtain the results (\ref{NJLYbeta}), (\ref{NJLYdim}), (\ref{NJLsigma2}) and (\ref{betabNJL}).


\bibliographystyle{ssg}
\bibliography{GN-GNY}

\end{document}